\documentclass[twocolumn]{jpsj3}

\usepackage{txfonts}
\def\m{\mathbf}\def\ua{\uparrow}\def\da{\downarrow}\def\ra{\rightarrow}

\def\ma{\mathrm}\def\si{\sigma}
\def\={\hspace{-2mm}&=&\hspace{-2mm}}

\def\etal{{\it et\ al.\ }}

\newcommand{\lsim}
 {\ \raise.35ex\hbox{$<$}\kern-0.75em\lower.5ex\hbox{$\sim$}\ }
\newcommand{\gsim}
 {\ \raise.35ex\hbox{$>$}\kern-0.75em\lower.5ex\hbox{$\sim$}\ }
\def\journal #1#2#3#4{#1 {\bf #2} (#4) #3}
\def\PR{Phys.\ Rev.}
\def\PRA{Phys.\ Rev.\ A}
\def\PRB{Phys.\ Rev.\ B}
\def\PRL{Phys.\ Rev.\ Lett.}

\def\JCP{J.\ Chem.\ Phys.}
\def\JLTP{J.~Low Temp.~Phys.}

\def\JPCM{J.\ Phys.\ Condens.\ Matter}

\def\JPSJ{J.\ Phys.\ Soc.\ Jpn.}

\def\RMP{Rev.\ Mod.\ Phys.}
\def\PTP{Prog.\ Theor.\ Phys.}
\def\ZP{Z.\ Phys.}

\def\EPJB{Eur.\ Phys.\ J.\ B}

\def\RPP{Rep.\ Prog.\ Phys.}
\def\PRep{Phys.\  Rep.}
\def\NP{Nat.\ Phys.}
\def\NJP{New\ J.\ Phys.}
\def\PProc{Phys.\ Proc.}
\title{
Variational Monte Carlo Study of Spin-Gapped Normal State and \\ 
BCS-BEC Crossover in Two-Dimensional Attractive Hubbard Model
} 
 
\author{
Shun \surname{Tamura}
\thanks{E-mail address: shun@cmpt-serv.phys.tohoku.ac.jp} and 
Hisatoshi {\sc Yokoyama} 
}
\inst{
\address{Department of Physics, Tohoku University, Sendai, 980-8578, Japan} 
}

\abst{
We study the properties of normal, superconducting (SC), and CDW states for 
an attractive Hubbard model on the square lattice, using a variational 
Monte Carlo method. 
In trial wave functions, we introduce an interspinon binding 
factor, indispensable for inducing a spin-gap transition in the normal 
state, in addition to the onsite attractive and intersite repulsive factors. 
It is found that, in the normal state, as the interaction strength $|U|/t$ 
increases, a first-order spin-gap transition arises at $|U_{\rm c}|\sim W$ 
($W$: bandwidth) from a Fermi liquid to a spin-gapped state, 
which is conductive as a result of the hopping of doublons. 
In the SC state, we confirm by the analysis of various quantities that 
the mechanism of superconductivity undergoes a smooth crossover 
at approximately $|U_{\ma{co}}|\sim |U_{\rm c}|$ from a BCS type to a Bose-Einstein 
condensation (BEC) type, as $|U|/t$ increases. 
For $|U|<|U_{\ma{co}}|$, quantities such as the condensation energy, 
a SC correlation function and the condensate fraction of onsite pairs 
exhibit the behavior of $\sim \exp(-t/|U|)$, as expected from the BCS theory.  
For $|U|>|U_{\ma{co}}|$, quantities such as the energy gain in the SC 
transition and superfluid stiffness, which is related to the cost of 
phase coherence, behave as $\sim t^2/|U|\propto T_{\rm c}$, as expected 
in a bosonic scheme. 
In this regime, SC transition is induced by a gain in kinetic 
energy, in contrast to the BCS theory.
We refer to the relevance to the pseudogap in cuprate superconductors. 
}

\kword{superconductivity, BCS-BEC crossover, pseudogap, superfluid stiffness, 
condensate fraction, CDW, attractive Hubbard model, square lattice, 
variational Monte Carlo method}
\begin{document}
\maketitle

\section{Introduction\label{sec:intro}}
%
Since the pioneering effort by Eagles,\cite{Eagles} many researchers have 
extensively and repeatedly addressed the transition of superconducting (SC) 
properties from a BCS type to a Bose-Einstein condensation (BEC) type as the 
strength of attractive potential between fermions is increased. 
Following early studies of the BCS-BEC crossover\cite{Leggett,Miyake} 
using continuum models with the superfluidity of $^3$He in mind, Nozi\`eres 
and Schmitt-Rink\cite{N-SR} showed by approximation that the SC properties 
smoothly evolve with the correlation strength in an attractive Hubbard model 
(AHM). 
Later, stimulated by the discovery of high-$T_{\rm c}$ cuprates with 
a small coherence length, numerous researchers have tackled 
AHM,\cite{Micnus} especially in two dimensions (2D); now, in connection 
with the evolution of pseudogaps as the doping rate $\delta$ decreases 
in the so-called underdoped regime,\cite{pseudogap} the problem of BCS-BEC 
crossover as a function of $\delta$ is a subject of 
urgency.\cite{cuprate,Levin} 
Entering this century, we have become capable of directly observing phenomena 
of crossover\cite{Regal,Zwierlein} and pseudogaps\cite{Gaebler,Feld} 
in traps of ultracold dilute alkali gases,\cite{Bloch,Giorgini} for which 
physical parameters can be artificially tuned. 
Recent experimental advances have brought hope of obtaining similar observations 
on optical lattices. 
\par 

In the above stream of research, AHM is one of the most important and basic 
lattice models for studying the evolution of SC properties according to 
the interaction strength $U/t$ ($U$: onsite interaction strength, 
$t$: hopping integral between nearest-neighbor sites). 
In early and later studies of AHM, mean-field-type\cite{Micnus} and 
diagrammatic\cite{N-SR,Bickers,Deisz,TPSC} approaches were used; 
although they successfully treated the weakly correlated regime, 
where the original BCS theory is basically valid, and developed 
a conceptual framework of the BCS-BEC crossover, more reliable methods 
remain necessary to establish properties in the intermediately 
(unitary) and strongly correlated (BEC) regimes. 
First, as an unbiased way, quantum Monte Carlo (QMC) calculations 
were implemented in the weakly and intermediately correlated regimes 
($|U|\lsim W$, $W$: bandwidth), because QMC is free from the 
negative-sign problem for AHM, but statistical fluctuation increases 
with increasing $|U|/t$ and system size. 
In 2D, SC transition of the Berezinskii-Kosterlitz-Thouless type was 
confirmed and an $n$-dependent phase diagram was discussed 
($n$: particle density).\cite{Moreo} 
Then, it was shown in the normal state ($T>T_{\rm c}$) for intermediate 
$|U|/t$'s that a thermal-activation-type behavior appears 
in magnetic susceptibility, but that the charge compressibility 
is almost $T$-independent.\cite{Randeria} 
This spin-gap behavior was corroborated by a peak split in the density 
of state.\cite{Singer} 
The dynamical mean field theory (DMFT), which becomes exact in infinite 
dimensions and is applicable to an arbitrary interaction strength, is 
another important approach to AHM. 
Early DMFT studies addressed normal branches without introducing 
SC orders at low temperatures ($T<T_{\rm c}$\cite{Keller} and 
$T=0$\cite{Capone}), and found that the normal state undergoes 
a first-order transition from a Fermi liquid to a gapped state 
at $|U|/W=1$-1.5, as $|U|/t$ increases. 
Later, various properties of the SC phase were 
calculated,\cite{Garg,Bauer,Koga} 
and the crossover was characterized by the SC gap and superfluid 
stiffness.\cite{Toschi}
\par 

Another effective approach to AHM is a many-body variation theory, 
which is applicable continuously in the entire range of correlation 
strengths and particle densities. 
In contrast to DMFT, the dimension and lattice form are realistically 
specified, and one can treat wave-number-dependent properties 
in low-lying states. 
Furthermore, since wave functions are explicitly given, this approach 
has advantages in forming a physical picture. 
Because an AHM of a bipartite lattice is mapped to a repulsive Hubbard 
model (RHM) by a canonical transformation,\cite{Dichtel,Shiba,Nagaoka} 
one can develop a theory 
relying on the knowledge of RHM. 
Thus, the well-known Gutzwiller wave function (GWF)\cite{GWF} became 
a primary trial function for the normal state; first, its properties 
were studied\cite{Medina} using the so-called Gutzwiller approximation 
(GA).\cite{GA} 
As known for RHM, although GWF itself is always metallic,\cite{YS1} 
additional GA induces a spurious metal-insulator (Brinkman-Rice) 
transition\cite{BR} at finite $|U_{\rm BR}|/t$ in finite 
lattice dimensions. 
For $|U|>|U_{\rm BR}|$ in AHM, all the particles tightly form onsite 
singlet pairs, and hopping completely ceases, so that the Brinkman-Rice 
transition remains a metal-insulator (Mott) transition also in AHM. 
Later, approximations similar to GA, which may be correct in infinite 
dimensions, have also been applied to the SC state\cite{Suzuki,Saito,Bunemann} 
to discuss the BCS-BEC crossover. 
However, to avoid the ambiguity of GA in realistic dimensions and 
to make use of the merits of the variation method, we need 
to accurately estimate variational expectation values. 
This claim is satisfied by a variational Monte Carlo (VMC) 
method,\cite{McMillan,Ceperley,YS1,Umrigar} which treats local 
correlation factors exactly without a minus sign problem. 
A decade ago, a VMC method was applied to a normal state in AHM 
to study a transition corresponding to the Mott transition in RHM 
by introducing a binding factor between adjacent antiparallel 
spinons.\cite{Y-PTP} 
For simplicity, we call a singly occupied site a spinon. 
However, the interpretation of the transition was incorrect on account 
of the limitation of treated system sizes and an insufficient analysis. 
Recently, VMC has been applied to solving problems with optical lattices 
in a confinement potential.\cite{Fujiwara}
\par

In this study, on the basis of VMC calculations of high precision 
for normal, SC and CDW states, we modify the previous results\cite{Y-PTP} 
and make features of the BCS-BEC crossover in AHM on the square lattice 
microscopically more clear. 
We mainly discuss the following points: 
(1) In both normal and SC states, a correlation between adjacent 
antiparallel spinons, in addition to the Gutzwiller correlation, is 
indispensable to qualitatively derive proper behavior.
(2) In the normal state, which underlies the SC state, a first-order 
phase transition occurs at $|U_{\ma{c}}|\sim W$ from a Fermi-liquid 
to a spin-gapped state. 
This transition is caused by the competition between the size of 
an antiparallel-spinon pair and the interpair distance, as in the case 
of Mott transitions in RHM.\cite{Miyagawa,boson} 
(3) The properties of SC noticeably change at approximately 
$|U_{\ma{co}}|\sim|U_{\rm c}|$, which are compared with those derived 
in a strongly correlated RHM for high-$T_{\rm c}$ cuprates. 
Part of the present result has been published before.\cite{Tamura}
\par

The rest of this paper is organized as follows: 
In \S\ref{sec:formulation}, we explain the model and method used 
in this study. 
In \S\ref{sec:normal}, we provide a discussion of the spin-gap 
transition arising in the normal state, and of the features 
in the spin-gapped regime. 
In \S\ref{sec:SC}, we consider a BCS-BEC crossover from various 
points of view. 
In \S\ref{sec:summary}, we briefly summarize our main results. 
\par

\section{Formulation\label{sec:formulation}} 
In \S\ref{sec:model}, we introduce AHM and briefly mention its relation 
to RHM. 
In \S\ref{sec:wf}, we discuss trial wave functions for normal, SC, 
and CDW phases. 
In \S\ref{sec:VMC}, we briefly explain the setup of VMC calculations 
in this work. 
\par

\subsection{Attractive Hubbard model\label{sec:model}}
We consider a single-band attractive Hubbard model ($U\leq0$) 
on a square lattice: 
\begin{equation}
  {\cal H}={\cal H}_t+{\cal H}_U 
= \sum_{{\bf k}\sigma}\varepsilon_{\bf k} 
                          c_{{\bf k}\sigma}^{\dag}c_{{\bf k}\sigma}
  +U\sum_{j}n_{j\uparrow}n_{j\downarrow}, 
\label{eq:model}
\end{equation}
%
where $n_{j\sigma}=c_{j\sigma}^{\dag}c_{j\sigma}$, $c_{j\sigma}$, and 
$c_{{\bf k}\sigma}$ are fermion annihilation operators in the Wannier 
and Bloch representations, respectively, and
\begin{equation}
  \varepsilon_{\bf k}=-2t(\cos k_x +\cos k_y). 
\label{eq:dispersion}
\end{equation}
We use the hopping integral $t$ and lattice constant as the units 
of energy and length, respectively. 
Because the lattice has a particle-hole symmetry at $n=N/N_{\rm s}=1$ 
($N$: number of particles, $N_{\rm s}$: number of lattice sites),
and properties at half filling are deduced from the results of 
RHM using corresponding wave functions,\cite{YTOT,YOT} as mentioned 
below, we mostly treat cases of $n<1$. 
The chemical potential term $-\zeta\sum_{j\sigma}n_{j\sigma}$ may be 
added to eq.~(\ref{eq:model}) to adjust particle density, if necessary. 
\par

In the following, we summarize the relation of AHM to RHM. 
The attractive Hubbard Hamiltonian eq.~(\ref{eq:model}) on a bipartite 
lattice satisfying the relation 
$\varepsilon_{\bf k}=-\varepsilon_{-{\bf k}+{\bf Q}}$ 
[here ${\bf Q}=(\pi,\pi)$] is mapped by the canonical 
transformation\cite{Dichtel,Shiba} 
\begin{equation}
c_{{\bf k}\uparrow}=\tilde{c}_{{\bf k}\uparrow}, \quad
c_{{\bf k}\downarrow}=\tilde{c}^\dag_{-{\bf k}+{\bf Q}\downarrow}
\label{eq:canonical}
\end{equation}
to RHM with constant shifts:
\begin{equation}
\tilde{\cal H}=\sum_{{\bf k}\sigma}\varepsilon_{\bf k}
 \tilde c_{{\bf k}\sigma}^\dag\tilde c_{{\bf k}\sigma} 
 +|U|\sum_j\tilde n_{j\uparrow}\tilde n_{j\downarrow}
 +UN\tilde n_\uparrow
 -h\sum_j\left(\tilde S^z_j+\frac{1}{2}\right), 
\end{equation}
where $\tilde n_{j\sigma}=\tilde c_{j\sigma}^\dag\tilde c_{j\sigma}$, 
$\tilde S^z_j=(\tilde n_{j\uparrow}-\tilde n_{j\downarrow})/2$ and 
$\tilde n_\sigma=\tilde N_\sigma/N$. 
A tilde denotes the representation transformed according 
to eq.~(\ref{eq:canonical}). 
The chemical potential $\zeta$ and $n$ in AHM are related to 
the effective magnetic field as $h=2\zeta$ and to the magnetization 
as $m=1-n$ in the $z$-direction in RHM, respectively. 
Therefore, unless the original AHM has a spin polarization ($m=0$), 
the particle density in the transformed RHM is always at half filling
($\tilde n=1$). 
Also, the order parameters of CDW and onsite singlet pairing 
defined as 
\begin{eqnarray}
O_{\rm CDW}&=&\frac{1}{N}\left|\sum_j e^{i{\bf Q}\cdot{\bf r}_j}
\langle n_{j\uparrow}+n_{j\downarrow}-1\rangle\right|, \\
O_{\rm SC}&=&\frac{1}{N}\sum_j
\langle c_{j\uparrow}^\dag c_{j\downarrow}^\dag\rangle \quad\mbox{or}\quad 
\frac{1}{N}\sum_j
\langle c_{j\downarrow}c_{j\uparrow}\rangle, 
\end{eqnarray}
in AHM are transformed into the forms of the $z$- and $xy$-components 
of the SDW order parameter:
\begin{eqnarray}
\tilde O_{\rm SDW}^z&=&\frac{1}{N}\left|\sum_j e^{i{\bf Q}\cdot{\bf r}_j}
\langle \tilde n_{j\uparrow}-\tilde n_{j\downarrow}\rangle\right|, \\
\tilde O_{\rm SDW}^\pm&=&\frac{1}{N}\sum_j
\langle\tilde c_{j\uparrow}^\dag\tilde c_{j\downarrow}\rangle 
\quad\mbox{or}\quad \frac{1}{N}\sum_j
\langle\tilde c_{j\downarrow}^\dag\tilde c_{j\uparrow}\rangle, 
\end{eqnarray}
respectively, in RHM.\cite{Nagaoka}
It is widely accepted that, at $T=0$, an antiferromagnetic (AFM) long-range 
order with equal magnitudes of $O_{\rm SDW}^\alpha$ for $\alpha=x,y,z$ 
arises in the half-filled RHM on the square lattice for arbitrary $U$ ($>0$), 
and that the AFM order in the $z$-direction is easily destroyed 
by a field applied in the $z$-direction $h$, whereas the AFM orders 
in the $xy$-plane survive. 
This implies that the ground state of AHM possesses a singlet pairing 
order for any $U$ and $n$ ($\zeta$), and simultaneously possesses a CDW 
order of the same magnitude at half filling $n=1$ ($\zeta=0$).\cite{Nagaoka}  
This argument was confirmed by direct calculations for AHM.\cite{Micnus} 
Although the above mapping holds unconditionally in exact treatments, 
when some approximation is applied, the validity of the mapping 
has to be verified individually for each specific treatment. 
\par 

\subsection{Trial wave functions\label{sec:wf}}
As a development of our previous study,\cite{Y-PTP} we apply a many-body 
variation theory to the Hamiltonian eq.~(\ref{eq:model}). 
As a trial wave function, a two-body Jastrow-type 
$\Psi={\cal P}\Phi_{\ma{MF}}$ was adopped,\cite{Jastrow} 
where $\Phi_{\ma{MF}}$ is a one-body (mean-field) wave function and 
${\cal P}$ is a many-body correlation (Jastrow) factor. 
\par

As the many-body part, we use the form  
${\cal P}={\cal P}_f{\cal P}_Q{\cal P}_{\rm G}$ in this work. 
The onsite (Gutzwiller) projector 
\begin{equation}
 {\cal P}_{\rm G}=\prod_{j}
         \left[1-\left(1-g\right)d_j\right], 
\label{eq:gp}
\end{equation}
with $d_j=n_{j\uparrow}n_{j\downarrow}$, is the most important. 
The variational parameter $g$ increases the number of doubly occupied 
sites (doublons), and ranges over $1\le g<\infty$ for $U\le 0$; 
in the limit of $g\rightarrow\infty$, singly occupied sites (spinons) 
are not allowed in a nonmagnetic case. 
If we put $\tilde g=1/g$, the properties of ${\cal P}_{\rm G}(\tilde g)$ 
for RHM are applicable to the present case ${\cal P}_{\rm G}(g)$.\cite{Y-PTP} 
\par

To explain the importance of a binding factor between the up and down spinons 
${\cal P}_Q$, it is convenient to refer to an effective Hamiltonian 
in the strong-correlation limit ($t/|U|\rightarrow 0$):\cite{Emery}
\begin{equation}
 {\cal H}_{\rm eff}=\frac{2t^2}{|U|}
 \sum_{<i,j>}\left[\left(-b^{\dag}_ib_j+\rho_i\rho_j+\si_i\si_j\right)
                   +\mbox{H.c.}-\frac{1}{2} \right], 
\label{eq:ste1}
\end{equation}
with
\begin{eqnarray}
 b_i=c_{i\ua}c_{i\da},\;\;
 \rho_i=\frac{1}{2}(n_{i\ua}+n_{i\da}-1),\;\;
 \si_i=\frac{1}{2}(n_{i\ua}-n_{i\da}). 
\label{eq:b}
\end{eqnarray}
The first term of eq.~(\ref{eq:ste1}) indicates the hopping of doublons. 
The second is a repulsive interaction between doublons [or empty sites 
(holons)] and an attractive interaction between a doublon and a holon 
in nearest-neighbor (NN) sites.
The third works as an AFM-Ising interaction. 
The expectation values of these terms can be reduced using antiparallel-spinon 
configurations in NN sites. 
To encourage such configurations, we introduce the attractive intersite 
correlation,\cite{Y-PTP} 
\begin{eqnarray} 
{\cal P}_Q\=\prod_j\left(1-\mu Q_j\right) 
\label{eq:PQ} \\
    Q_j\=s^{\ua}_{j}\prod_{\tau}
         (1-s^{\da}_{j+\tau})+s^{\da}_{j}\prod_{\tau}(1-s^{\ua}_{j+\tau})
\label{eq:Qj}
\end{eqnarray}
where $s_j^{\sigma}=n_{j\sigma}(1-n_{j-\sigma})$ (spinon projector), 
and $\tau$ runs over NN sites of the site $j$. 
In ${\cal P}_Q$, the parameter $\mu$ ($0\le\mu\le 1$) controls the strength 
of binding between NN antiparallel spinons; 
for $\mu=0$, spinons are free of binding, while in the limit 
$\mu\rightarrow 1$, antiparallel spinons are necessarily paired as nearest 
neighbors. 
As we will see later, ${\cal P}_Q$ is indispensable for a spin-gap 
transition\cite{Y-PTP} and a proper description of the SC state. 
In fact, ${\cal P}_Q$ is the canonical transformation 
through eq.~(\ref{eq:canonical}) of the doublon-holon binding factor 
often used to describe Mott transitions in RHM.\cite{Kaplan,YS3} 
A Mott transition in RHM corresponds to a spin-gap transition in AHM, 
as we will see in \S\ref{sec:normal}.  
Since $Q_j$ is a spin-dependent projector, the so-called spin 
contamination\cite{contamination} arises in the wave function, namely, 
$\Psi$ deviates from an eigenstate of ${\bf S}^2=(\sum_j{\bf S}_j)^2$. 
However, in this case, the expectation values of $\langle{\bf S}^2\rangle$ 
estimated using a VMC method are as small as 0.15 (2) in the SC 
(normal) state for $N=200$-$300$ at its maximum at $U\sim W$. 
Because these values, particularly of the SC state, are much smaller than 
those of the AFM state, the spin contamination is considered to have little 
influence on the results. 
\par

As a factor supplementary to ${\cal P}_Q$, a repulsive correlation 
suited to eq.~(\ref{eq:ste1}) should be considered. 
As a simple one, we check a repulsive factor between NN doublons: 
\begin{eqnarray} 
 {\cal P}_f=\prod_j
 \left[1-fd_j\left(1-\prod_{\tau}\bar d_{j+\tau}\right)\right], 
\label{eq:pf} 
\end{eqnarray} 
where $f$ ($0\le f\le 1$) is a parameter,  
$\bar d_j=1-d_j$, and $\tau$ runs over NN sites of the site $j$. 
The projector ${\cal P}_f$ reduces the weight of configurations with 
adjacent doublons by $1-f$; for $f\rightarrow 0$, the effect of 
${\cal P}_f$ vanishes, and for $f\rightarrow 1$, a doublon cannot sit 
in a NN site of another doublon. 
\par

Now, we turn to the one-body part $\Phi$ of the wave function. 
For a normal state, we adopt the Fermi sea $\Phi_{\rm F}$. 
Since general features of $\Psi_{\rm N}={\cal P}\Phi_{\rm F}$ with $f=0$ were 
studied in a previous paper,\cite{Y-PTP} here we focus on the properties 
of the transition arising at $U\sim W$, which was regarded as a Mott 
transition.\cite{Y-PTP} 
\par

It is known that the BCS state $\Phi_{\rm BCS}$ can deal with 
the BCS-BEC crossover in some degree;\cite{Leggett,N-SR} it is natural 
to employ $\Phi_{\rm BCS}$ for a SC state: 
\begin{eqnarray}
 \Phi_{\rm BCS}=\left(\sum_{\m{k}}a_{\m{k}}
c_{\m{k}\ua}^{\dag}c_{-\m{k}\da}^{\dag} \right)^{N/2}|0\rangle.
\label{eq:BCS}
\end{eqnarray}
where the particle number is fixed and 
\begin{eqnarray}
 a_{\m{k}}=\frac{v_{\m{k}}}{u_{\m{k}}}=
 \frac{\Delta_{\rm P}}{\varepsilon_{\m{k}}-\bar\zeta+
 \sqrt{(\varepsilon_{\m{k}}-\bar\zeta)^2+\Delta_{\rm P}^2}}, 
\label{eq:ak}
\end{eqnarray}
with $\Delta_{\rm P}$ and $\bar\zeta$ being variational parameters 
corresponding to the SC gap $\Delta_{\rm SC}$ and chemical potential 
$\zeta$, respectively, in the weakly correlated limit, and 
$$
u_{\bf k}^2\ (v_{\bf k}^2)=\frac{1}{2}\left[
1+(-)\frac{\varepsilon_{\rm k}-\bar\zeta}
{\sqrt{\left(\varepsilon_{\bf k}-\bar\zeta\right)^2
                 +\Delta_{\rm p}^2}}\right]. 
$$ 
For $\Delta_{\rm P}\rightarrow 0$, $\Phi_{\rm BCS}$ is reduced to 
$\Phi_{\rm F}$. 
Here, we assume $\Delta_{\rm P}$ to be a homogeneous $s$ wave on account 
of the attractive contact potential. 
For RHM, a form similar to eq.~(\ref{eq:ak}) with a $d_{x^2-y^2}$-wave pair 
potential was studied,\cite{YTOT,Y12} where, as $\delta$ decreases, what 
$\Delta_{\rm P}$ means deviates from the SC gap, and represents 
a pseudogap.\cite{ZGRS,Paramekanti} 
In contrast, in the present case, $\Delta_{\rm P}$ seems to reflect 
the magnitude of $T_{\rm c}$ for any $|U|/t$, except for $n\rightarrow 1$, 
for which $T_{\rm c}$ is considered to vanish owing to the CDW order.
The correlated SC function $\Psi_{\rm sc}={\cal P}\Phi_{\rm BCS}$ is 
mapped using eq.~(\ref{eq:canonical}) to a projected AFM wave function 
ordered in the $x$-$y$ plane at $n=1$. 
\par 

In addition, we check a CDW wave function for $n\sim 1$: 
\begin{equation}
 \Phi_{\rm CDW}=\prod_{{\bf k},\sigma}
\left(-\alpha_{\bf k}c_{{\bf k}\sigma}^\dag
       +\beta_{\bf k}c_{{\bf k}+{\bf Q}\sigma}^\dag\right)|0\rangle, 
\label{eq:CDW}
\end{equation} 
where the ${\bf k}$ sum is taken in the Fermi sea, $\m{Q}=(\pi,\pi)$, and
\begin{equation} 
 \alpha_{\bf k}\ (\beta_{\bf k})=
\sqrt{\frac{1}{2}\left(1-(+)\frac{\varepsilon_{\bf k}}
 {\sqrt{\varepsilon_{\bf k}^2+\Delta_{\rm c}^2}}\right)}\ , 
\label{eq:CDWcoeff}
\end{equation}
with $\Delta_{\rm c}$ being a parameter corresponding to the CDW gap. 
$\Psi_{\rm CDW}={\cal P}\Phi_{\rm CDW}$ is mapped through 
eq.~(\ref{eq:canonical}) to a projected AFM wave function ordered 
in the $z$-direction at $n=1$. 
Because the AFM order is isotropic in RHM, $\Psi_{\rm SC}$ and 
$\Psi_{\rm CDW}$ should yield identical results at half filling. 
\par

\subsection{Variational Monte Carlo method\label{sec:VMC}}
In estimating variational expectation values with respect to $\Psi$ 
discussed in \S\ref{sec:wf}, we use a VMC 
method,\cite{McMillan,Ceperley,YS1,Umrigar} which gives virtually 
exact values for finite but relatively large systems.
Since the number of variational parameters is not large, we execute 
rounds of linear optimization for each parameter with the other 
parameters fixed until the parameters as well as the energy converge 
(typically 3-5 rounds) with $2.5\times 10^5$ particle configurations 
generated through a Metropolis algorithm. 
After the convergence, we continue to execute additional $20$ to $30$ 
rounds of iteration with successively renewed configuration sets.
We determine the optimized values by averaging the data obtained in the 
additional rounds; in averaging, we exclude scattered data beyond the range 
of twice the standard deviation.
Thus, the optimal value is an average of substantially more than several 
million samples. 
Physical quantities are computed with the optimized parameters thus 
obtained with $2.5\times 10^5$ samples. 
\par

We use systems of $L\times L$ sites of up to $L=32$ for $\Psi_{\rm N}$ 
and $L=24$ for $\Psi_{\rm SC}$ with the periodic-antiperiodic 
boundary conditions to reduce level degeneracy. 
We choose the particle densities to satisfy the closed-shell condition, 
and mainly study $n=0.25$ (0.26), 0.5, and 0.75. 
\par

\section{Spin-Gap Transition in Normal State\label{sec:normal}}
As mentioned in \S\ref{sec:model}, the ground state of AHM is SC
for any $U/t$ and $n$ (and CDW at $n=1$). 
Therefore, the normal state, we address in this section, is not 
the ground state of eq.~(\ref{eq:model}). 
The significance to study $\Psi_{\rm N}$ is not only in a passive sense 
that a normal state appears when the SC state is destroyed 
by, e.g., magnetic field or impurities, but in that $\Psi_{\rm N}$ 
underlies $\Psi_{\rm SC}$, just as a SC transition is understood 
by the instability of the Fermi sphere against an infinitesimal 
attractive interaction in the BCS theory. 
Namely, normal states are deeply involved in the mechanism of 
SC transitions. 
\par

In \S\ref{sec:energy}, we show the improvement in energy by introducing 
${\cal P}_Q$ and ${\cal P}_f$, and the energy gains using the SC 
and CDW states. 
In \S\ref{sec:spingap}, we confirm the existence of a spin-gap transition 
in the normal state. 
In \S\ref{sec:picture}, we consider the mechanism of the spin-gap transition 
and other properties. 
\par

\subsection{Energy improvement\label{sec:energy}}
First, we briefly look at the energy improvement by the projection 
factors ${\cal P}_Q$ and ${\cal P}_f$ in the normal, SC, and CDW states. 
\par

\begin{table}[htbp]
\vspace{-0.0cm} 
\caption{
Comparison of energy per site $E/t$ among three kinds of projection factors 
${\cal P}_{\rm G}$, ${\cal P}_{\cal Q}\equiv{\cal P}_Q {\cal P}_{\rm G}$, and 
${\cal P}_{\cal F}\equiv{\cal P}_f {\cal P}_Q {\cal P}_{\rm G}$ in normal 
and SC states. 
A system of $n=0.5$ and $L=12$ is used; the tendency is the same 
for other $n$'s. 
The last digit in each section includes some statistical errors. 
}
\begin{center}
\begin{tabular}{r|l|l|l|l} 
\hline
\multicolumn{1}{c|}{$|U|/t$} & \multicolumn{1}{c|}{1} & 
\multicolumn{1}{c|}{3}       & \multicolumn{1}{c|}{7} & 
\multicolumn{1}{c}{10} \\
\hline \hline
${\cal P}_{\rm G}\ \Phi_{\ma{F}}$ & 
 $-1.37513$ & $-1.54791$ & $-2.0677$ & $-2.6215$ \\ \hline
${\cal P}_{\cal Q}\ \Phi_{\ma{F}}$ &
 $-1.37536$ & $-1.55099$ & $-2.10435$& $-2.7417$ \\ \hline
${\cal P}_{\cal F}\ \Phi_{\ma{F}}$ & 
 $-1.37536$ & $-1.55097$ & $-2.10440$& $-2.7416$ \\ \hline\hline
${\cal P}_{\rm G}\ \Phi_{\ma{BCS}}$ & 
 $-1.37531$ & $-1.55686$ & $-2.17815$ & $-2.80500$ \\ \hline
${\cal P}_{\cal Q}\ \Phi_{\ma{BCS}}$ & 
 $-1.375535$& $-1.55908$ & $-2.18444$ & $-2.81162$ \\ \hline
${\cal P}_{\cal F}\ \Phi_{\ma{BCS}}$ & 
 $-1.375534$& $-1.55918$ & $-2.18681$ & $-2.81436$ \\ \hline
\end{tabular}
\label{table:energy}
\vspace{-0.4cm} 
\end{center}
\end{table}
As studied in detail in ref.~\citen{Y-PTP}, the variational energy in 
the normal state is considerably improved by ${\cal P}_Q$ on that of GWF, 
especially for large $|U|/t$'s (Table \ref{table:energy}). 
Moreover, it is known that the phase transition discussed in \S\ref{sec:spingap} 
does not arise without ${\cal P}_Q$.\cite{YS1,Y-PTP} 
Thus, the factor ${\cal P}_Q$ is indispensable to appropriately describe 
the normal state. 
These aspects of ${\cal P}_Q$ correspond to those of the doublon-holon 
(D-H) binding factor in RHM.\cite{YS3,YOT} 
On the other hand, the improvement by ${\cal P}_f$ 
on ${\cal P}_Q{\cal P}_{\rm G}\Phi$ is almost imperceptible for any $|U|/t$ 
as shown in Table \ref{table:energy}. 
The optimized parameter $f$ is nearly zero, namely, 
${\cal P}_f$ scarcely modifies the wave function. 
\par

In the SC state, the improvement in $E/t$ by ${\cal P}_Q$ 
on ${\cal P}_{\rm G}\Phi_{\rm BCS}$ is not as large as that 
for the normal state. 
This is because the effect of binding between up and down spinons, i.e., 
the effect of singlet pair creation, is already included in 
$\Phi_{\rm BCS}$ to some extent, as the one-body AFM state has some 
D-H binding effect for RHM.\cite{YTOT} 
Further energy reduction by ${\cal P}_f$ is again negligible for 
small $|U|/t$'s and remains relatively small in magnitude for larger $|U|/t$'s, 
as compared with the energy reduction by ${\cal P}_Q$. 
The magnitude of energy reduction by ${\cal P}_f$ is similarly small 
for $n=0.25$ and 0.75. 
\par

Since we find that a short-range repulsive factor ${\cal P}_f$ produces 
only negligible effects in all the cases we treat, we omit ${\cal P}_f$ 
and use the form 
${\cal P}={\cal P}_{\cal Q}\equiv{\cal P}_Q{\cal P}_{\rm G}$ 
as the many-body factor in $\Psi$ in the rest of this paper, unless 
otherwise specified. 
\par

\begin{figure}[htbp]
\begin{center}
\includegraphics[width=8.5cm,clip]{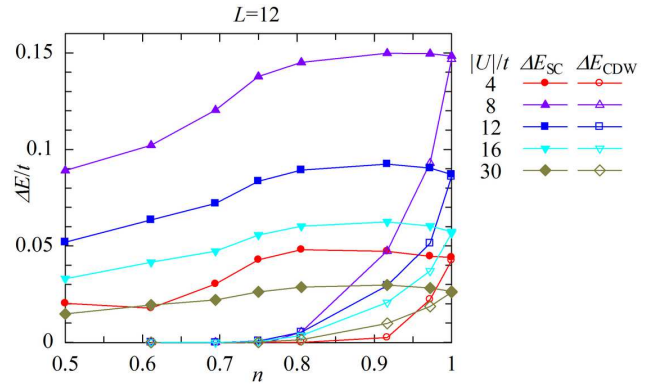} 
\end{center}
\caption{(Color online)
Particle-density dependences of the energy gains by $\Psi_{\rm SC}$ 
and $\Psi_{\rm CDW}$ are plotted for five values of $|U|/t$. 
For small $|U|/t$'s, the system size dependence (fluctuation) is large 
owing to the large coherence length. 
}
\vspace{-0.3cm} 
\label{fig:cdw}
\end{figure}
Finally, we compare the energy gains using the SC and CDW states: 
\begin{equation}
\Delta E_{\rm SC}\ (\Delta E_{\rm CDW})=
E_{\rm N}-E_{\rm SC}\ (E_{\rm CDW}), 
\label{eq:Egain}
\end{equation}
where $E_{\rm N}$, $E_{\rm SC}$, 
and $E_{\rm CDW}$ are the optimized energies per site for $\Psi_{\rm N}$, 
$\Psi_{\rm SC}$ and $\Psi_{\rm CDW}$, respectively. 
Figure \ref{fig:cdw} shows the $n$ dependences of $\Delta E_{\rm SC}$ and 
$\Delta E_{\rm CDW}$ for large $n$'s. 
At half filling ($n=1$), the SC and CDW states are degenerate, but this 
degeneracy is immediately lifted for $n<1$. 
$\Delta E_{\rm CDW}$ rapidly deteriorates and vanishes as $n$ decreases, 
whereas $\Delta E_{\rm SC}$ preserves appreciable values for high densities 
and gradually decays until $n=0$ (not shown). 
This feature of $\Delta E$ coincides with what we discussed for 
the canonical transformation in \S\ref{sec:model}.\cite{Micnus} 
\par

In the remainder of this section, we will concentrate on $\Psi_{\rm N}$. 
\par

\subsection{Spin-gap transition\label{sec:spingap}}
In a previous VMC study using ${\cal P}_{\cal Q}\Phi_{\rm F}$,\cite{Y-PTP} 
a transition was detected with systems up to $L=12$ 
at $U=U_{\rm c}\sim 9t$. 
However, this transition was misinterpreted as a continuous metal-insulator 
transition. 
In this subsection and the next, we study the features of this transition 
more carefully. 
\par

\begin{figure}[htbp]
\begin{center}
\includegraphics[width=8.0cm,clip]{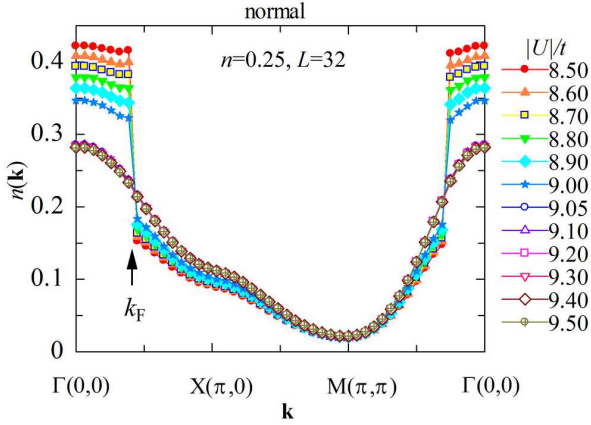}
\end{center}
\caption{(Color online)
Momentum distribution function for some values of $|U|/t$ near transition 
point $|U_{\ma{c}}|/t$ along path $(0,0)$-$(\pi,0)$-$(\pi,\pi)$-$(0,0)$. 
The discontinuity at ${\bf k}_{\rm F}$ indicated by an arrow is used 
to estimate $Z$ shown in Fig.~\ref{fig:Z}. 
}
\vspace{-0.6cm} 
\label{fig:nk}
\end{figure}
First, we confirm the existence of a transition. 
In Fig.~\ref{fig:nk}, we plot the momentum distribution function 
\begin{equation}
 n(\m{k})=\frac{1}{2}\sum_\sigma
     \left\langle c_{\m{k}\sigma}^{\dag}c_{\m{k}\sigma} \right\rangle
\label{eq:nk}
\end{equation}
for $|U|/t\sim 9$ and $n=0.25$ ($L=32$). 
For $|U|/t\le 9.0$, $n({\bf k})$ has discontinuities on the $\Gamma$-X and 
$\Gamma$-M segments, indicating that a Fermi surface exists and 
the state is a Fermi liquid. 
On the other hand, the discontinuity suddenly vanishes for $|U|/t\ge 9.05$, 
and $n({\bf k})$ becomes a smooth function of ${\bf k}$. 
It follows a certain gap opens and the state becomes a non-Fermi liquid 
for $|U|>|U_{\rm c}|$, with $9.0<|U_{\rm c}|/t<9.05$ in this case. 
Through similar analyses, we found $|U_{\rm c}|/t\sim 0.875$ (0.83) 
for $n=0.195$ (0.121) for $L=32$; $|U_{\rm c}|/t$ tends to gradually 
decrease with $n$.\cite{note-Uc} 
Thus, a transition from a Fermi liquid to a non-Fermi liquid certainly 
exists, as found in our previous study.\cite{Y-PTP} 
According to similar analyses for $L=24$ and 28 and $n\sim 0.25$, 
the system-size dependence of $|U_{\rm c}|/t$ is very small at these values 
of $L$, but $|U_{\rm c}|/t$ tends to increase slightly as $L$ increases. 
Such a feature is analogous to those of the Mott transitions 
in RHM induced by D-H binding factors.\cite{YOT,Miyagawa,boson} 
\par 

\begin{figure}[htbp]
\begin{center}
\includegraphics[width=7.5cm,clip]{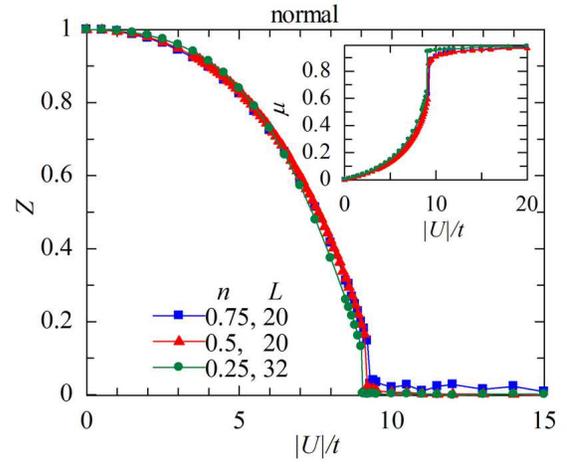}
\end{center}
\caption{(Color online) 
Quasiparticle renormalization factor as function of $|U|/t$ for three 
particle densities, estimated from discontinuities of $n({\bf k})$ 
in Fig.~\ref{fig:nk} and other data. 
The inset shows the optimized spinon-binding parameter as a function 
of $|U|/t$ near $U=U_{\rm c}$ for the same systems as that in the main panel.  
}
\label{fig:Z}
\end{figure}
Next, we check the order of this transition. 
In Fig.~\ref{fig:Z}, the quasiparticle renormalization factor $Z$ is shown 
vs $U/t$; $Z$ is obtained using $Z=n(k_\ma{F}-0)-n(k_\ma{F}+0)$ 
on the $\Gamma$-X segment, where the values of $n(k)$ 
at $k\rightarrow k_{\rm F}\pm 0$ are estimated using third-order 
least-squares fits of the data for $k<k_{\rm F}$ and $k>k_{\rm F}$, 
respectively. 
There exist clear discontinuities in $Z$ at $U=U_{\rm c}$. 
The optimized spinon-binding parameter $\mu$ plotted in the inset 
of Fig.~\ref{fig:Z} also exhibits a large jump at $U=U_{\rm c}$. 
In fact, other physical quantities show a similar discontinuous behavior. 
Thus, we may safely conclude that this transition is not a continuous 
transition but a first-order phase transition. 
The reason why the previous study could not find the correct transition 
order is that the discontinuous behavior of $n({\bf k})$ manifests itself 
only for $L\gsim 18$. 
The critical value $|U_{\rm c}|/t$ only slightly depends on $n$. 
\par

Now, we consider the features of $\Psi_{\rm N}$ in the non-Fermi-liquid 
regime $|U|>|U_{\rm c}|$. 
As shown in the inset of Fig.~\ref{fig:Z}, the spinon-binding parameter 
$\mu$ approaches unity, suggesting that almost all up and down spinons 
are paired as singlets. 
In discussing gap formation, the small-$|{\bf q}|$ behavior of the charge 
(density) and spin structure factors
\begin{eqnarray}
 N(\m{q})\=\frac{1}{N}
      \sum_{j,l}e^{-i\m{q}\cdot\m{r}_l}\langle n_j n_{j+l}\rangle-n^2 
\label{eq:Nq} \\
 S(\m{q})\=\frac{1}{N}
      \sum_{j,l}e^{-i\m{q}\cdot\m{r}_l}\langle S^z_j S^z_{j+l}\rangle 
\label{eq:Sq}
\end{eqnarray}
provide us with useful information. 
Assuming that the lowest excitation occurs at ${\bf q}={\bf 0}$, the energy 
gap in the spin sector between the ground state $|\Psi_0\rangle$ 
and the first excited state $|\Psi_\m{q}\rangle$ is given 
by the single-mode approximation (SMA)\cite{SMA} as 
\begin{eqnarray}
 \Delta_S
  \=\frac{\langle\Psi_{\bf q}|({\cal H}-E_0)|\Psi_{\bf q}\rangle}
         {\langle\Psi_{\bf q}|\Psi_{\bf q}\rangle} 
   =\frac{\langle\Psi_0|[S_{-{\bf q}},[{\cal H},S_{\bf q}]]|\Psi_0\rangle}
         {\langle \Psi_{\bf q}|\Psi_{\bf q} \rangle} \nonumber \\
  \=-\frac{1}{8}\lim_{q\ra0}\frac{q^2}{S(\m{q})}K
\label{eq:SMA}
\end{eqnarray}
where $K$ denotes the kinetic energy, 
$|\Psi_\m{q}\rangle=S_\m{q}|\Psi_0\rangle$, 
and 
\begin{equation}
S_\m{q} = \frac{1}{\sqrt{N}}\sum_j e^{i\m{q}\cdot\m{r}_j}S^z_j\ . 
\end{equation}
From eq.~(\ref{eq:SMA}), we find that $\Delta_S$ vanishes 
if $S(\m{q})\propto q$ for $q\rightarrow 0$, whereas $\Delta_S$ becomes 
finite, if $S(\m{q})\propto q^2$ for $q\rightarrow 0$. 
The charge (density) gap $\Delta_N$ can be similarly treated. 
\par

\begin{figure}[htbp]
\vspace{-0.1cm} 
\begin{center}
\includegraphics[width=8.0cm,clip]{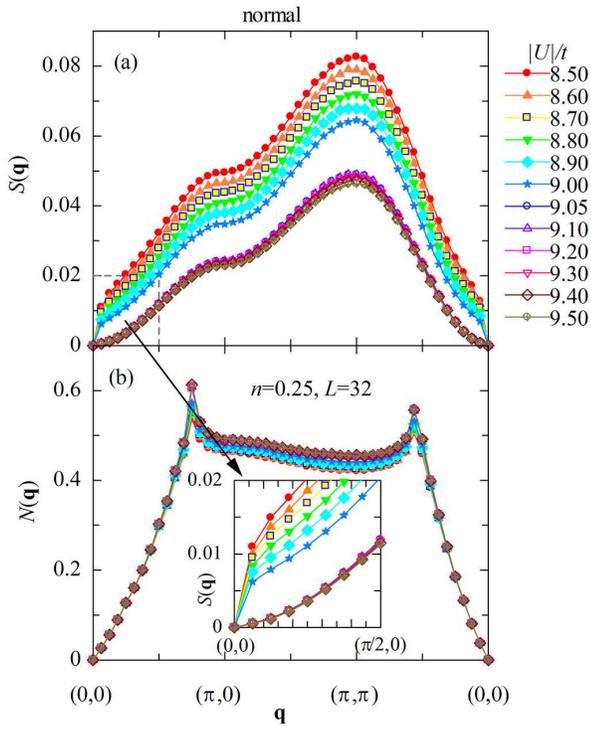}
\end{center}
\caption{(Color online)
(a) Spin and (b) charge (density) structure factors along path 
$(0,0)$-$(\pi,0)$-$(\pi,\pi)$-$(0,0)$ for some values of $|U|/t$ 
near $|U_{\rm c}|/t$ ($9.0<|U_{\rm c}|/t<9.05$). 
The inset in (b) is a magnification of $S({\bf q})$ near 
${\bf q}\sim{\bf 0}$ on the segment $(0,0)$-$(\pi,0)$. 
The tendency is the same for other values of $n$.
}
\vspace{-0.1cm} 
\label{fig:SqNq_normal}
\end{figure}
\begin{figure}[htbp]
\vspace{-0.0cm} 
\begin{center}
\includegraphics[width=6.5cm,clip]{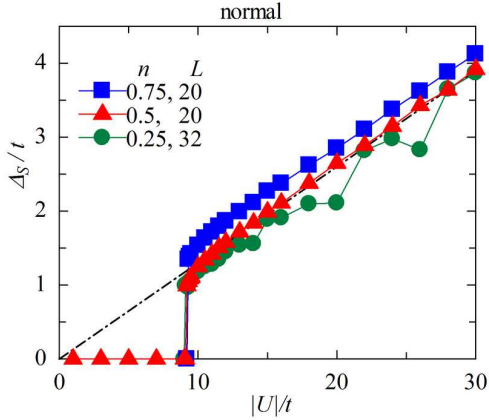}
\end{center}
\caption{(Color online)
Spin gaps estimated by single-mode approximation for three particle 
densities as functions of $|U|/t$. 
The dash-dotted line is a visual guide for $n=0.5$. 
The spin-gap transition occurs at $|U_\m{c}|/t\sim 9$. 
}
\vspace{-0.3cm} 
\label{fig:spingap}
\end{figure}

In Fig.~\ref{fig:SqNq_normal}, we show $S(\m{q})$ and $N(\m{q})$ 
for some values of $|U|/t$ near $U=U_{\rm c}$. 
In the vicinity of $\m{q}=\m{0}$, as $|U|/t$ increases, $S(\m{q})$ 
abruptly changes its behavior from linear to quadratic at $U=U_{\rm c}$, 
as shown in the inset of Fig.~\ref{fig:SqNq_normal}(b). 
Thus, it is very likely that the spin gap is generated in the non-Fermi 
liquid regime. 
In Fig.~\ref{fig:spingap}, we plot the spin gap estimated using 
eq.~(\ref{eq:SMA}) for the segment of $(0,0)$-$(\pi,0)$ of $S(\m{q})$; 
the magnitude of $\Delta_S$ is proportional to $|U|$ 
($\Delta_S\sim 0.13|U|$ for $n=0.5$), and depends on $n$ only weakly. 
In contrast, the behavior of $N(\m{q})$ shown 
in Fig.~\ref{fig:SqNq_normal}(b) is almost unchanged including the 
$2k_{\rm F}$ anomaly, if $|U|/t$ varies, and remains linear in $q$ 
for $q\rightarrow 0$ for $|U|>|U_{\rm c}|$.
Thus, low-energy properties with respect to the charge degree of 
freedom are unlikely to be affected by this transition; 
regarding the charge excitation, $\Psi_{\rm N}$ remains gapless 
and conductivity is preserved in the spin-gapped regime. 
This may be the first realization of a conductive spin-gapped normal 
state in the variation theory. 
\par

\subsection{Picture of transition and spin-gapped state
\label{sec:picture}}
%
\begin{figure}[htbp]
\begin{center}
\includegraphics[width=7.0cm,clip]{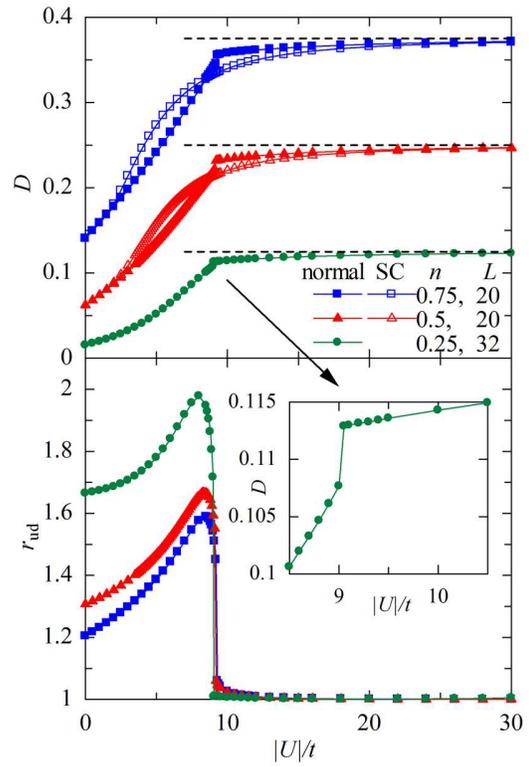}
\end{center}
\caption{(Color online)
(a) Doublon density as function of $|U|/t$ for $\Psi_{\rm N}$ and 
$\Psi_{\rm SC}$. 
The dashed lines indicate the maximum $D$, i.e., $n/2$. 
The inset in (b) shows the magnification of $D$ ($\Psi_{\rm N}$) for 
$n=0.25$ near the transition point, at which $D$ displays a small jump.
(b) Average distances from up spinon to its nearest down spinon 
for three particle densities as functions of $|U|/t$. 
}
\vspace{-0.3cm} 
\label{fig:dist_ud}
\end{figure}
To deepen our understanding of the above spin-gap transition, let us look at 
some other quantities. 
Figure \ref{fig:dist_ud}(a) shows the doublon density  
\begin{equation}
D=\frac{1}{N_{\ma{s}}}\sum_j\langle b^\dag_jb_j\rangle.
\end{equation}
As $|U|/t$ increases, $D$ increases in the Fermi-liquid state 
owing to the attractive correlation of ${\cal P}_{\rm G}$, 
but it reaches almost its full value ($n/2$) at $U=U_{\rm c}$. 
The main panel of Fig.~\ref{fig:dist_ud}(b) shows the average distance 
from an up (down) spinon to its nearest down (up) spinon $r_{\rm ud}$. 
Here, we measure distance $r$ by the stepwise (so-called 
Manhattan) metric. 
As $|U|/t$ increases in a small-$|U|/t$ regime, $r_{\rm ud}$ increases 
because the densities of up and down spinons decrease owing to 
doublon formation, and the binding correlation of ${\cal P}_Q$ 
is still weak, as in the inset of Fig.~\ref{fig:Z}. 
However, $r_{\rm ud}$ abruptly drops when $U$ approaches $U_{\rm c}$, 
and converges to unity for $|U|>|U_{\rm c}|$, because an up spinon and 
a down spinon are tightly bound within NN sites ($\mu\rightarrow 1$). 
Consequently, for $|U|>|U_{\rm c}|$, almost all particles form onsite 
pairs, and even if a doublon resolves into spinons, they remain 
an adjacent pair and do not itinerate as isolated spinons. 
\par 

\begin{figure}[htbp]
\begin{center}
\includegraphics[width=8.5cm,clip]{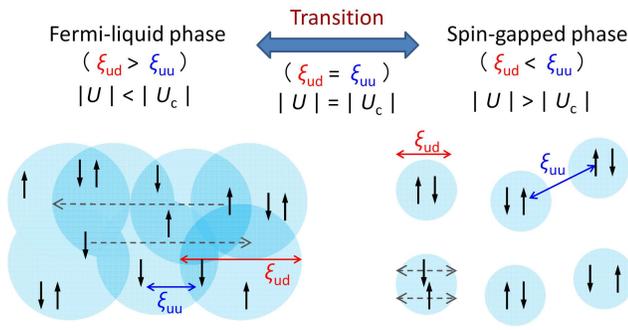}
\end{center}
\caption{(Color online)
Illustration of mechanism of spin-gap transition. 
The up and down arrows denote up and down spinons, respectively. 
Typical configurations are shown for the two phases. 
Although in each pair domain drawn with a pale circle, at least one up 
spinon and one down spinon must exist, excess spinons can still move around 
independently, slipping out of their original domain, as indicated 
by the long dashed arrows in the Fermi-liquid phase. 
In the spin-gapped phase, a spinon cannot itinerate independently of 
the partner spinon out of the pair domain. 
}
\vspace{-0.3cm} 
\label{fig:up-down}
\end{figure}
Thus, we notice that this spin-gap transition can be understood 
in parallel with a recently proposed picture of Mott transitions 
owing to the D-H binding.\cite{Miyagawa,boson} 
Here, we postulate that antiparallel spinon pairs with a pair domain 
of size $\xi_{\rm ud}$ are created by the attractive correlation 
of ${\cal P}_Q$. 
We can appropriately define this binding length $\xi_{\rm ud}$ and 
also the minimum distance from a spinon to its nearest parallel 
spinon $\xi_{\rm uu}$ as 
\begin{eqnarray} 
 \xi_{\rm ud}\=r_{\rm ud}+\sigma_{\rm ud}, 
\label{eq:xiud} \\ 
 \xi_{\rm uu}\=r_{\rm uu}-\sigma_{\rm uu}, 
\label{eq:xiuu}
\end{eqnarray}
where $r_{\rm uu}$ is the average distance from an up (down) spinon 
to its nearest up (down) spinon, and $\sigma_{\rm ud}$ and 
$\sigma_{\rm uu}$ are the standard deviations of $r_{\rm ud}$ and 
$r_{\rm uu}$, respectively. 
In the spin-gapped phase, the relation $\xi_{\rm ud}<\xi_{\rm uu}$ holds, 
indicating that the domains of pairs do not usually overlap, at least, 
not in sequence. 
Consequently, almost all pairs are isolated and an up spinon and a down spinon 
are confined within $\xi_{\rm ud}$, resulting in singlet pairs 
of small lengths with finite excitation gaps. 
In contrast, in the Fermi-liquid phase, $\xi_{\rm ud}$ becomes longer 
than $\xi_{\rm uu}$, indicating that the domains of spinon pairs overlap 
with one another. 
Then, an up spinon in a pair can exchange a partner down spinon 
with a down spinon in an adjacent pair. 
As a result, an up spinon and a down spinon can move independently by exchanging 
their partner, as shown in long arrows in Fig.~\ref{fig:up-down}, 
and definite singlet pairs cannot be specified. 
Thus, as $|U|/t$ is varied, a spin-gap transition takes place 
when $\xi_{\rm ud}$ becomes equivalent to $\xi_{\rm uu}$, which is 
expected to be a monotonically increasing function of $|U|/t$. 
\par

\begin{figure}[htbp]
\vspace{-0.3cm} 
\begin{center}
\includegraphics[width=7.0cm,clip]{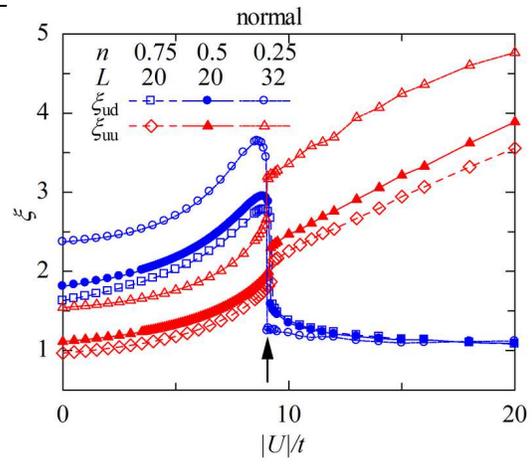}
\end{center}
\vspace{-0.3cm} 
\caption{(Color online)
The binding length of up and down spinon pairs, $\xi_{\rm ud}$, and 
the minimum distance between spinon pairs, $\xi_{\rm uu}$, defined 
by eqs.~(\ref{eq:xiud}) and (\ref{eq:xiuu}), respectively, 
are plotted as functions of $U/t$ for three densities. 
An arrow indicates the spin-gap transition point, the $n$ dependence of 
which is small.
}
\label{fig:xi_normal}
\end{figure}

Figure \ref{fig:xi_normal} shows $\xi_{\rm ud}$ and $\xi_{\rm uu}$ 
estimated from the VMC results as functions of $|U|/t$ for three 
particle densities. 
As expected from Fig.~\ref{fig:dist_ud}(b), $\xi_{\rm ud}$ abruptly 
drops at $U=U_{\rm c}$, whereas $\xi_{\rm ud}$ monotonically increases 
as $|U|/t$ increases with a small jump at the transition point. 
As a result, $\xi_{\rm ud}$ and $\xi_{\rm uu}$ intersect each other 
at $U=U_{\rm c}$ for any $n$. 
Thus, the scheme illustrated in Fig.~\ref{fig:up-down} is justified 
to some extent. 
\par

\begin{figure}[htbp]
\begin{center}
\includegraphics[width=7.0cm,clip]{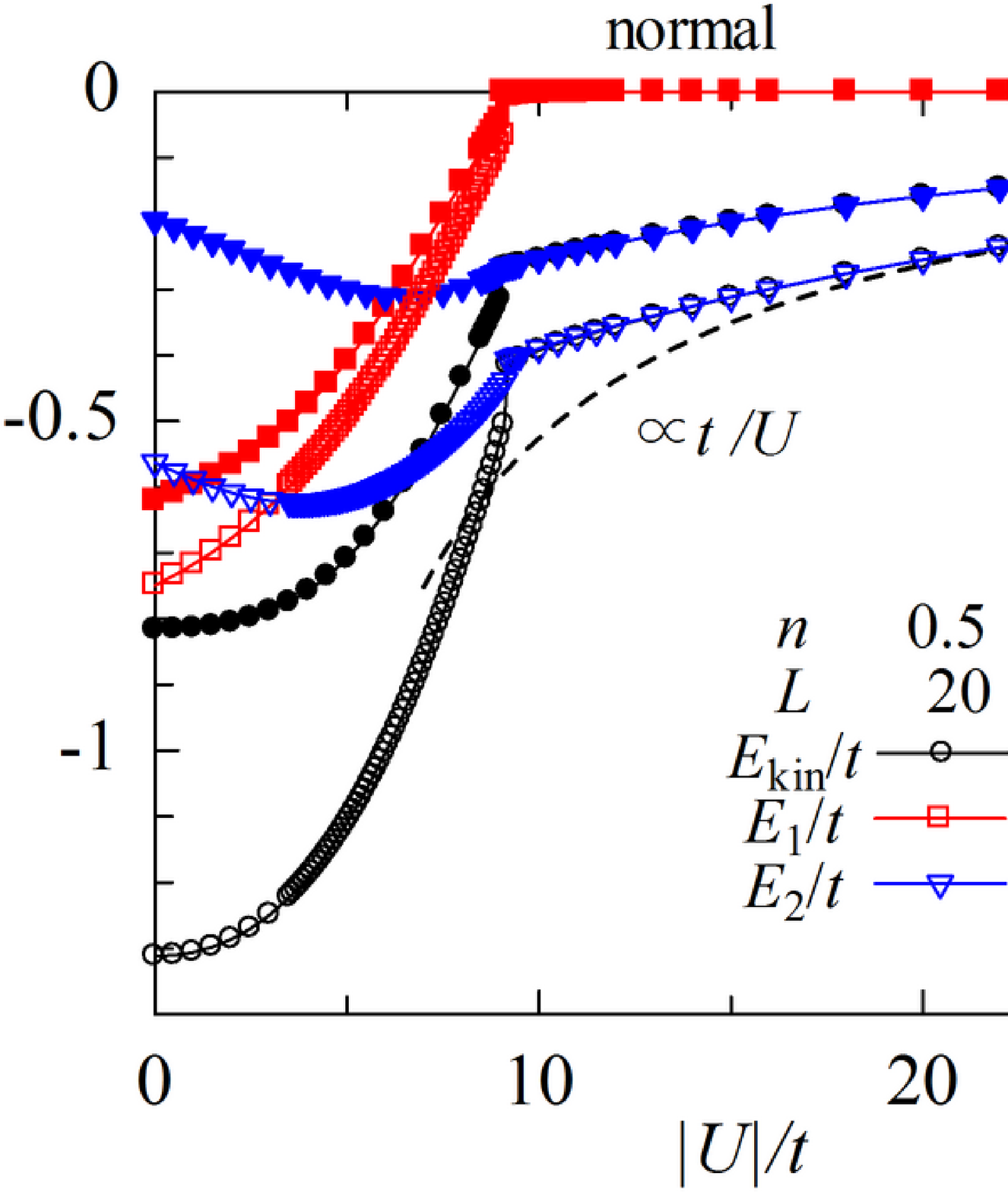}
\includegraphics[width=6.5cm,clip]{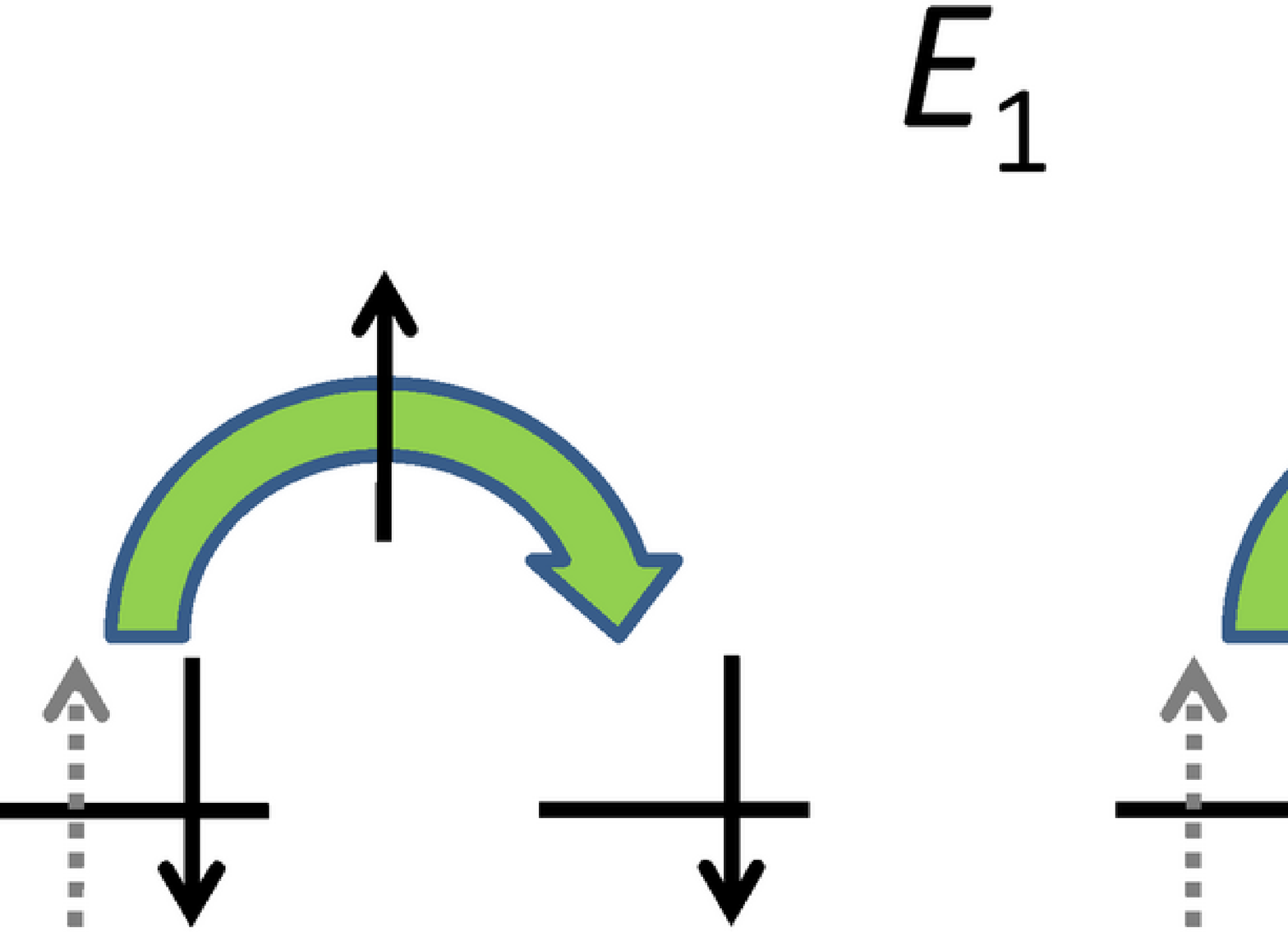}
\end{center}
\vspace{-0.3cm} 
\caption{(Color online)
In the main panel, the kinetic energy $E_\ma{kin}$ and its components 
$E_1$ and $E_2$ are depicted as functions of correlation strength 
for two particle densities. 
$E_1$ ($E_2$) is the contribution of hopping processes that do not 
(do) change $D$, as shown in the illustration in the lower part. 
The dashed line for $E_2/t$ is a guide proportional to $-t/|U|$ 
($|U|>|U_{\rm c}|$). 
}
\label{fig:E1E2}
\end{figure}
Finally, we discuss the itinerancy of particles. 
To this end, it is convenient to decompose the kinetic energy $E_{\ma{kin}}$ 
into two parts ($E_\ma{kin}=E_1+E_2$), namely, the contribution 
of the hopping processes that do not (do) change the number 
of doublons $E_1$ ($E_2$),\cite{Tocchio} 
as shown in the lower part of Fig. \ref{fig:E1E2}. 
In the main panel of Fig. \ref{fig:E1E2}, $E_1$, $E_2$, and $E_{\ma{kin}}$ 
are depicted as functions of $|U|/t$. 
For $|U|<|U_{\rm c}|$, both $E_1$ and $E_2$ contribute to $E_{\ma{kin}}$ 
because isolated spinons are independently mobile, whereas 
in the spin-gapped phase, $E_1$ almost vanishes ($E_{\ma{kin}}\sim E_2$) 
for any particle density. 
In this case, the independent motion of a spinon not accompanied 
by an antiparallel spinon is strongly suppressed. 
On the other hand, the contribution of the dissociation of a doublon 
into a spinon pair and their reunion, $E_2$, remains appreciable 
and is proportional to $-t^2/|U|$ for large $|U|/t$'s.  
This point is in sharp contrast to a feature of the Brinkman-Rice 
transition\cite{BR} derived using the Gutzwiller approximation;\cite{GA} 
in this case, the motion of particles is completely prohibited 
for $|U|>|U_{\rm BR}|=8|E(U=0)|$,\cite{Medina} so that the state becomes 
insulating, and a charge (Mott) gap opens. 
\par

\begin{figure}[htbp]
\begin{center}
\includegraphics[width=7.5cm,clip]{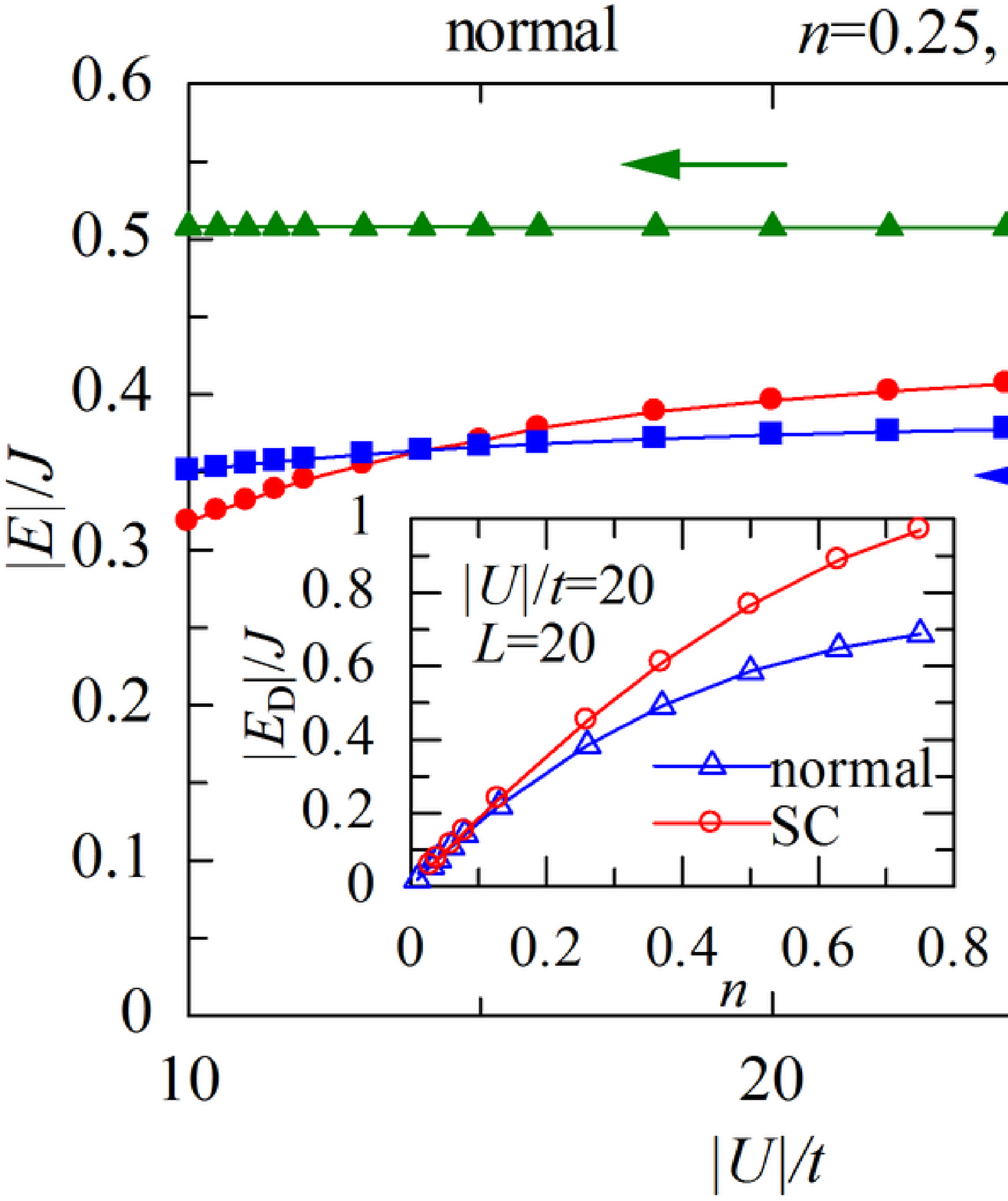}
\end{center}
\vspace{0.0cm} 
\begin{center}
\includegraphics[width=6.5cm,clip]{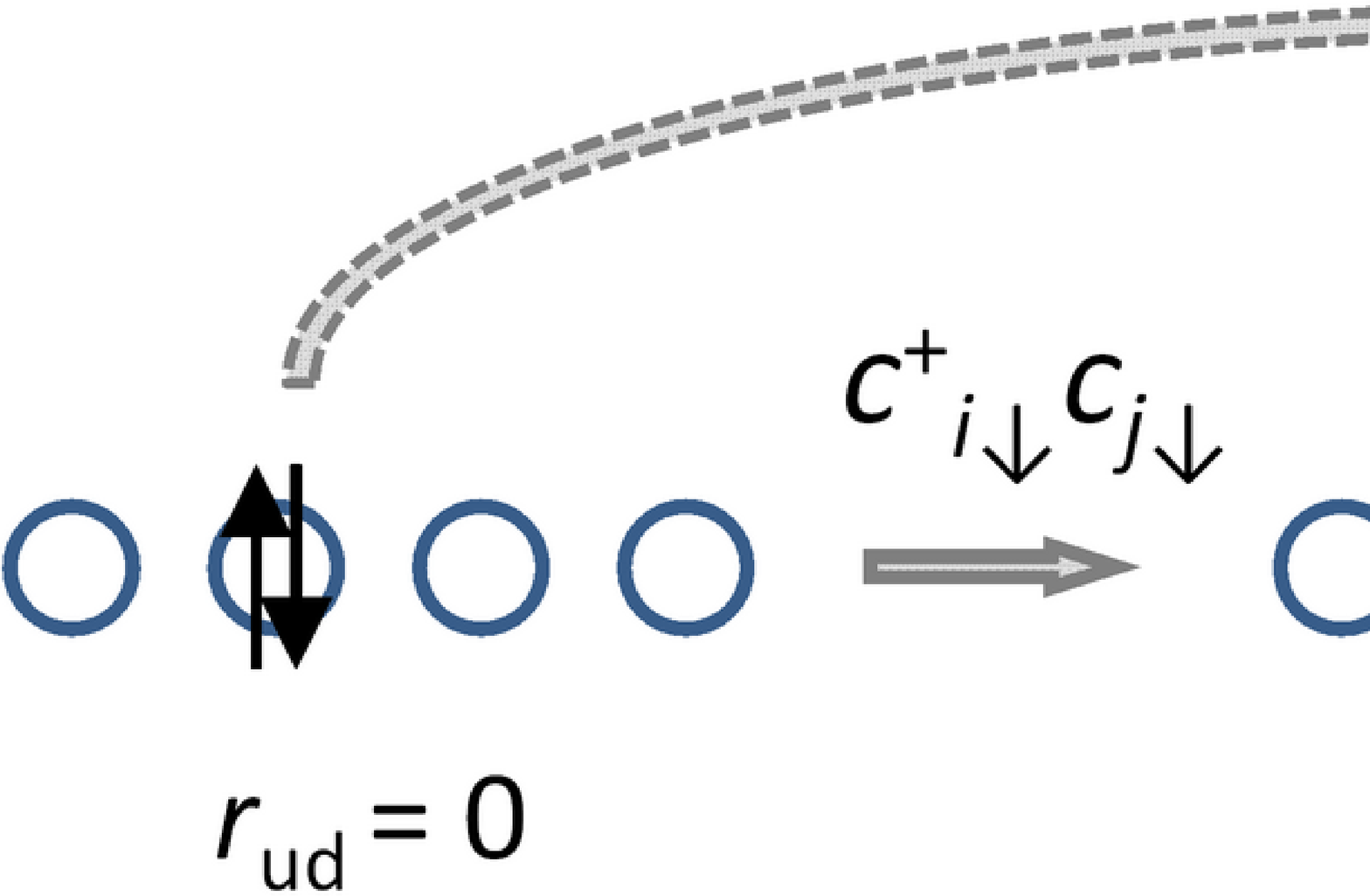}
\end{center}
\vspace{-0.3cm} 
\caption{(Color online) 
In the main panel, kinetic energy and two elements of 
$\langle {\cal H}_{\rm eff}\rangle$ [eqs.~(\ref{eq:ED}) and
(\ref{eq:Erhosigma})] 
normalized by $J=2t^2/U$ are plotted in the BEC regime as functions 
of $|U|/t$. 
Illustrated in the lower part is the relation between a single hopping 
process in ${\cal H}$ and a doublon hopping process in ${\cal H}_{\rm eff}$. 
The inset, discussed in \S\ref{sec:gain}, shows a comparison of the doublon 
hopping $E_{\rm D}$ between the normal and SC states as a function of particle 
density. 
}
\label{fig:Dh} 
\end{figure}
The above feature of $E_1=0$ ($|U|>|U_{\rm c}|$) for any $n$ is distinct 
from that of the D-H binding state for strongly correlated RHM, 
where $E_1$ is $n$-dependent ($\propto 1-n$) near half filling.\cite{Y12}  
Thus, in doped Mott insulators in RHM, doped holes or particles play 
a role of carriers, and bound D-H pairs are localized and not involved 
in conduction. 
On the other hand, in AHM, the motion of a doublon, which is composed of 
two single particle-hopping processes as shown at the bottom 
of Fig.~\ref{fig:Dh}, contributes to the kinetic energy. 
To check this, we calculate the doublon hopping and diagonal terms in
eq.~(\ref{eq:ste1}), namely, 
\begin{eqnarray}
\frac{E_{\ma{D}}}{J}\=
\frac{1}{N_{\rm s}}\sum_{j,\tau}\langle b_{j+\tau}^{\dag}b_j\rangle, 
\label{eq:ED}
\\
\frac{E_{\rho\sigma}}{J}\=
\frac{1}{N_{\rm s}}\sum_{j,\tau}\langle \rho_{j+\tau}\rho_j+
 \sigma_{j+\tau}\sigma_j\rangle, 
\label{eq:Erhosigma}
\end{eqnarray} 
with $J=2t^2/|U|$, in the original Hilbert space of eq.~(\ref{eq:model}). 
In the main panel of Fig.~\ref{fig:Dh}, we compare them with 
the single hopping contribution ($E_{\ma{kin}}\sim E_2$) for large $|U|/t$'s. 
In addition to the large constant contribution of $E_{\rho\sigma}/J$, 
the doublon hopping ($E_{\rm D}$) has an appreciable magnitude, 
indicating the possibility of transport. 
Thus, the normal state $\Psi_{\rm N}$ is conductive for any values 
of $|U|/t$ and $n$ ($\ne 1$).\cite{note-kappa}
\par

\section{Crossover of Superconducting Properties\label{sec:SC}}
In \S\ref{sec:gain}, we discuss the BCS-BEC crossover in the light of 
energy gain in the SC transition and of chemical potential, and show 
that the SC transition in the BEC regime is induced by the kinetic-energy 
gain. 
In \S\ref{sec:Ps}, we discuss quantities that characterize the 
BCS and BEC regimes. 
In \S\ref{sec:coherence}, we roughly estimate the coherence length 
and interpair distance, thereby giving an intuitive picture 
of the crossover.  
\par

\subsection{Energy gain and kinetic-energy-driven transition
\label{sec:gain}}
\begin{figure}[htbp]
\vspace{0.3cm} 
\begin{center}
\includegraphics[width=7.0cm,clip]{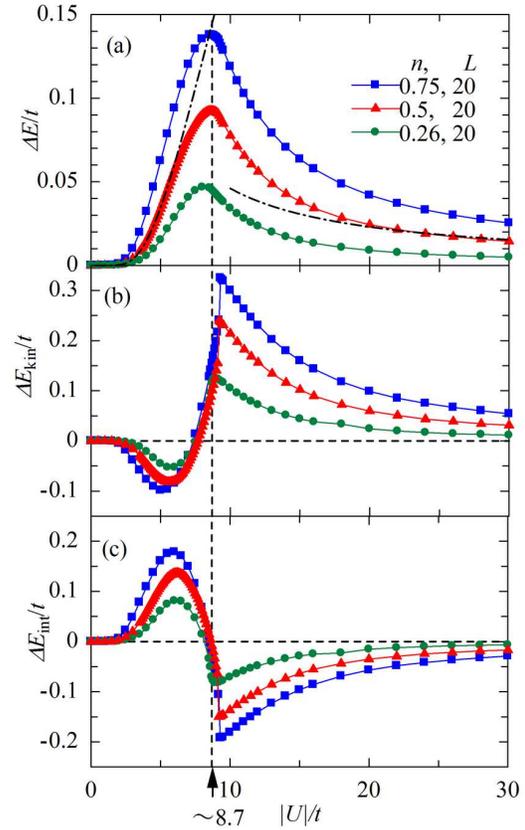}
\end{center}
\caption{(Color online) 
(a) Energy differences between normal and SC states as functions of $|U|/t$ 
for three particle densities. 
The maximum point is $|U_\ma{co}|/t\sim 8.7$ for these $n$'s, as indicated 
by a dashed line. 
The dash-dotted lines for $n=0.5$ are visual guides for the two limit 
$|U|/t\rightarrow 0$ [$\propto \exp(-\alpha t/|U|)$] and 
$\rightarrow\infty$ [$\propto \beta t/|U|$] with $\alpha$ and $\beta$ 
being constants. 
The kinetic and interaction components of $\Delta E$ are drawn in
(b) and (c), respectively.
}
\vspace{-0.5cm} 
\label{fig:cond_ene}
\end{figure}

First, we discuss the BCS-BEC crossover from the point of view of
the energy difference per site between the normal ($\Psi_{\rm N}$) 
and SC ($\Psi_{\rm SC}$) states $\Delta E$ $(\ge 0)$ defined 
in eq.~(\ref{eq:Egain}). 
In Fig.~\ref{fig:cond_ene}(a), the $|U|/t$ dependence of $\Delta E/t$ is 
shown; $\Delta E/t$ increases as $\sim\exp(-t/U)$ corresponding 
to the BCS theory for small $|U|/t$'s, reaches a maximum 
at $|U|=|U_{\rm co}|\sim 8.7t$, and then decreases for $|U|>|U_{\rm co}|$ 
as $\sim t/|U|$. 
As we will see later, various properties of SC actually exhibit qualitative 
changes at approximately this $|U_{\rm co}|/t$ from a BCS type to a BEC type. 
Note that normal-state properties are deeply involved in the 
crossover;\cite{note-co} 
$|U_{\rm co}|/t$ is affected by the spin-gap transition point 
$|U_{\rm c}|/t$ in $\Psi_{\rm N}$, where $E_{\rm N}$ exhibits a cusp, 
resulting in $U_{\rm co}\sim U_{\rm c}$. 
Recall that the normal state $\Psi_{\rm N}$, underlying $\Psi_{\rm SC}$, 
is a Fermi liquid for $|U|<|U_{\rm c}|$, but $\Psi_{\rm N}$ becomes 
a spin-gapped state in the absence of a Fermi surface, as shown in 
Fig.~\ref{fig:nk} for $|U|/t\ge 9.05$. 
Namely, for $|U|>|U_{\rm c}|$, a SC transition cannot be interpreted 
by the instability of the Fermi surface against an attractive interaction. 
In this relation, $\Delta E$ means the condensation energy 
for $|U|/t\sim 0$ according to the BCS theory, but $\Delta E$ probably 
deviates from the condensation energy observed experimentally 
for $|U|\gsim|U_{\rm co}|$, as in the case of high-$T_{\rm c}$ 
cuprates.\cite{Y12} 
\par

It is important to analyze $\Delta E$ into the kinetic part 
$\Delta E_{\rm kin}$ and the interaction part $\Delta E_{\rm int}$ 
($\Delta E=\Delta E_{\rm kin}+\Delta E_{\rm int}$). 
In the BCS theory, a SC transition is induced by lowering $E_{\rm int}$ 
at the cost of smaller loss in $E_{\rm kin}$. 
On the other hand, it is known in a large-$U/t$ regime of RHM that 
a SC transition occurs by reducing $E_{\rm kin}$ with a loss 
in $E_{\rm int}$.\cite{YTOT,Y12} 
In the latter case, the low-frequency sum rule of optical conductivity 
$\sigma_1(\omega)$\cite{Tinkham} should be broken, namely, high-frequency 
excitations in $\sigma_1(\omega)$ arise, because the sum of 
$\sigma_1(\omega)$ is proportional to $-E_{\rm kin}$\cite{Maldague} 
on a square lattice. 
In Figs.~\ref{fig:cond_ene}(b) and \ref{fig:cond_ene}(c), we show 
$\Delta E_{\rm kin}$ and $\Delta E_{\rm int}$, respectively, 
for AHM. 
For $|U|\lsim|U_{\rm co}|$, $E_{\rm int}$ ($E_{\rm kin}$) has a gain 
(loss) by the SC transition in accordance with the BCS theory. 
Meanwhile, for $|U|\gsim|U_{\rm co}|$, the situation is reverse; 
the SC transition is driven by a gain in kinetic energy. 
Correspondingly, the hopping of doublons (carriers) $|E_{\rm D}|$ 
becomes more enhanced in $\Psi_{\rm SC}$ than in $\Psi_{\rm N}$ 
in the BEC regime, as shown in the inset of Fig.~\ref{fig:Dh}. 
Kinetic-energy-driven (SC or magnetic) transitions may be 
rather general in strongly correlated systems.\cite{Nagaoka-Ferro,YTOT,Y12}
As mentioned previously, this reversal of driven force will be experimentally 
found if $\sigma_1(\omega)$ is accurately measured in cold-atom 
systems. 
In fact, similar results were reached for AHM in infinite dimensions 
by DMFT.\cite{Toschi,Kyung} 
\par 

\begin{figure}[htbp]
\vspace{0.0cm} 
\begin{center}
\includegraphics[width=8.5cm,clip]{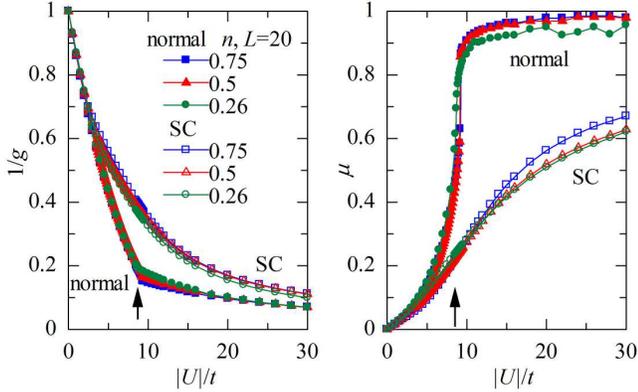}
\end{center}
\caption{(Color online) 
Optimized variational parameters, 
(a) Gutzwiller (onsite) factor $g$, and 
(b) NN antiparallel-spinon factor $\mu$ for $\Psi_{\rm N}$ and 
$\Psi_{\rm SC}$ are shown as functions of $|U|/t$. 
The symbols are common in (a) and (b). 
The arrows indicate the spin-gap transition points in $\Psi_{\rm N}$. 
}
\vspace{-0.5cm} 
\label{fig:g_mu}
\end{figure}

Let us consider this SC transition in the light of the variational 
parameters. 
At the level of one-body wave function, $\Phi_{\rm BCS}$ improves 
the energy over $\Phi_{\rm F}$ by creating onsite Cooper pairs 
through the pair potential $\Delta_{\rm P}$ (Fig.~\ref{fig:delta_var}),
in accordance with the BCS theory. 
Therefore, in the BCS regime, the number of doublons is expected to be more 
in the SC state than in the normal state. 
This is actually shown in Fig.~\ref{fig:dist_ud}(a). 
Since the increase in $D$ hinders the motion of particles, the kinetic 
energy is suppressed in the SC state, as in Fig.~\ref{fig:cond_ene}(b). 
As $|U|$ increases, however, $D$ of $\Psi_{\rm N}$ increases more rapidly 
and surpasses that of $\Psi_{\rm SC}$ at $|U_{\rm co}|$. 
This reversal is brought about mainly by the correlation factor ${\cal P}$. 
In Fig.~\ref{fig:g_mu}, we compare the optimized parameters in ${\cal P}$ 
between the normal and SC states. 
The onsite attractive factor $g$ is certainly larger in $\Psi_{\rm N}$, 
especially near $U=U_{\rm co}$. 
The antiparallel-spinon binding factor $\mu$ mainly works for the 
suppression of overgrown $D$ by $g$ in the BEC regime in order 
to gain $E_{\rm kin}$. 
\par

\begin{figure}[htbp]
\vspace{0.0cm} 
\begin{center}
\includegraphics[width=6.5cm,clip]{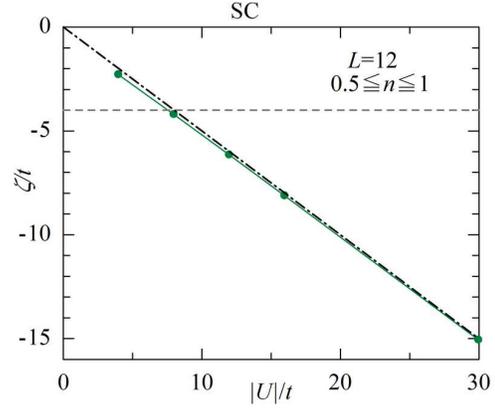}
\end{center}
\caption{(Color online) 
The chemical potential estimated from $E/t$ at higher particle densities 
($0.5\le n\le 1.0$) is shown by dots as a function of $|U|/t$. 
The dash-dotted line denotes the value at half filling: $\zeta=U/2$.
The dashed line indicates the band bottom 
$\varepsilon_{\rm L}$ ($-4t$). 
}
\vspace{-0.5cm} 
\label{fig:cp}
\end{figure}

In a one-body framework, when the chemical potential ($\zeta$) is 
situated in an energy band, low-energy excitation in the SC phase is 
described by Bogoliubov's quasiparticles, whereas when $\zeta$ becomes 
lower than the band bottom $\varepsilon_{\rm L}$, the statistics 
of the system becomes bosonic. 
Thus, the BCS-BEC crossover point is roughly estimated using 
$\zeta=\varepsilon_{\rm L}$. 
We estimate $\zeta$ for $\Psi_{\rm SC}$ from 
$\zeta=\partial E_{\rm SC}/\partial n$ (strictly finite differences). 
Within statistical error, $E_{\rm SC}$ is almost a linear 
function of $n$ for $0.5<n<1.0$, so that the $\zeta$ obtained 
in this range becomes independent of $n$. 
In Fig.~\ref{fig:cp}, we plot the thus-estimated $\zeta$ as a function of 
$|U|/t$ with the value at half filling i.e., $\zeta=U/2t$. 
$\zeta$ reaches the band bottom $\varepsilon_{\rm L}=-4t$ 
at $|U|/t\sim 7.9$. 
The behavior of $\zeta$ here is consistent with those obtained 
by DMFT.\cite{Garg,Bauer}
\par
\begin{figure}[htbp]
\vspace{0.0cm} 
\begin{center}
\includegraphics[width=7.5cm,clip]{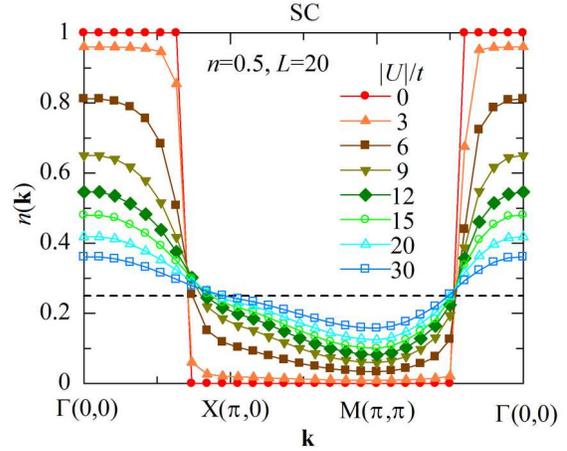}
\end{center}
\caption{(Color online) 
Evolution of momentum distribution function for $\Psi_{\rm SC}$ 
as $|U|/t$ varies. 
The dashed line indicates $|U|/t=\infty$. 
}
\vspace{-0.5cm} 
\label{fig:nk_sc}
\end{figure}
Finally, let us look at the momentum distribution function 
for $\Psi_{\rm SC}$ (Fig.~\ref{fig:nk_sc}). 
As $|U|/t$ increases, a step-function form of $|U|/t=0$ changes 
to a BCS type ($v_{\bf k}^2$) for small $|U|/t$ and is then smoothly 
modified through $|U_{\rm co}|/t$ to a constant in the BEC limit
($|U|/t=\infty$). 
Such evolution of $n({\bf k})$ has already been observed in an experiment 
on an ultracold gas in a trap;\cite{Regal-nk} we hope for similar experiments 
on optical lattices. 
\par

\subsection{Pair correlation function and helicity modulus
\label{sec:Ps}}
As mentioned in \S\ref{sec:intro}, a previous study of Toschi 
\etal\cite{Toschi}\ 
using DMFT 
argued that appropriate quantities that trace the strength of SC 
($T_{\rm c}$) in the BCS and BEC regimes are the gap 
$\Delta_{\rm SC}\sim\langle c^\dag_\uparrow c^\dag_\downarrow\rangle$ and 
the superfluid stiffness $D_{\rm s}$, respectively. 
$\Delta_{\rm SC}$ indicates the cost of creating a Cooper pair, while 
$D_{\rm s}$ characterizes the cost of realizeing phase coherence. 
In this subsection, we start with the quantities corresponding to 
$\Delta_{\rm SC}$ and $D_{\rm s}$. 
\par

\begin{figure}[htbp]
\begin{center}
\includegraphics[width=7.5cm,clip]{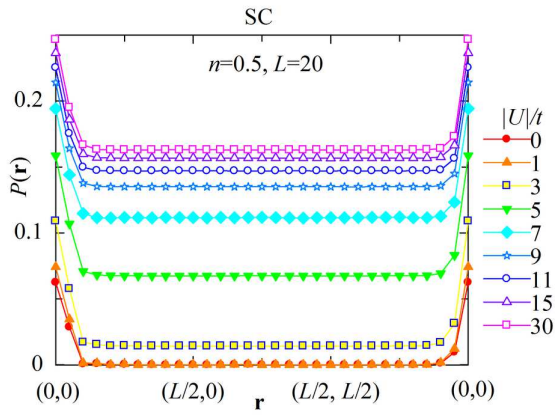}
\end{center}
\caption{(Color online) 
The onsite superconducting correlation function is drawn along 
a path of ${\bf r}_\ell$, (0,0)-$(L/2,0)$-$(L/2,L/2)$-(0,0), 
for various values of $|U|/t$. 
For other values of $n$ and $L$, the behavior is basically the same. 
}
\vspace{-0.1cm} 
\label{fig:P(r)}
\end{figure}
\begin{figure}[htbp]
\begin{center}
\includegraphics[width=8.0cm,clip]{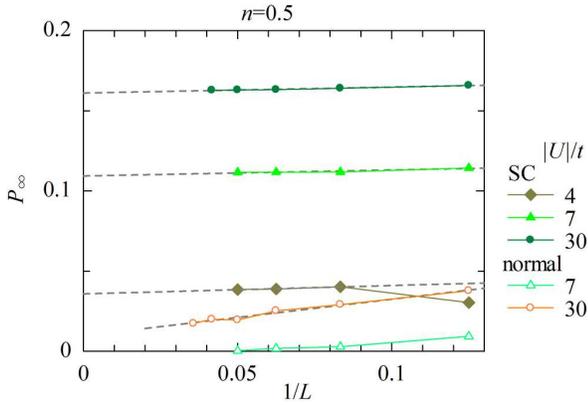}
\end{center}
\caption{(Color online) 
System size dependence of $P_\infty$ for three values of $|U|/t$ 
at $n=0.5$. 
In the cases of $L$ not satisfying the closed-shell condition at $n=0.5$, 
we adopt the average of $n=0.5\pm\delta$ with the smallest $\delta$ 
satisfying the condition. 
As visual guides, we show the first-order least-squares fits for $L\rightarrow\infty$ 
with dashed lines. 
Solid (open) symbols represent the data of $\Psi_{\rm SC}$ 
($\Psi_{\rm N}$). 
$\Psi_{\rm N}$ is bound to vanish for $L\rightarrow\infty$. 
}
\vspace{-0.5cm} 
\label{fig:P_infty_size}
\end{figure}
\begin{figure}[htbp]
\begin{center}
\includegraphics[width=8.5cm,clip]{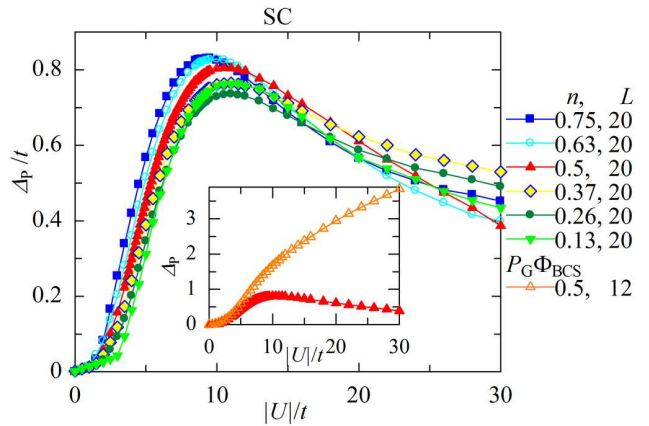}
\end{center}
\caption{(Color online)
Optimized gap parameter $\Delta_{\rm P}$ for several $n$'s as function of 
$|U|/t$. 
The dash-dotted line is a visual guide of $\propto \exp(-t/|U|)$.
The inset shows a comparison of the optimized values of $\Delta_{\rm P}$ between 
${\cal P}_Q{\cal P}_{\rm G}\Phi_{\rm BCS}$ and 
${\cal P}_{\rm G}\Phi_{\rm BCS}$ for $n=0.5$. 
}
\vspace{-0.3cm} 
\label{fig:delta_var} 
\end{figure}

As an appropriate index of off-diagonal-long-range order (ODLRO) 
in the present scheme, we consider a SC correlation function 
of onsite pairing,\cite{noteP} defined as 
\begin{eqnarray}
 P(\m{r}_\ell)=\frac{1}{N_{\rm s}}\sum_{j}
                   \langle b_{j}^{\dag}b_{j+\ell} \rangle. 
\label{eq:P}
\end{eqnarray}
%
The magnitude of ODLRO is given by the long-distance value 
of $P(\m{r}_\ell)$, i.e., 
$P_\infty=\lim_{|\m{r}_\ell|\ra \infty}P(\m{r}_\ell)\sim\Delta_{\rm SC}^2$. 
In the present VMC calculations with finite systems, we must check 
the ${\bf r_\ell}$ dependence of $P(\m{r}_\ell)$. 
In Fig.~\ref{fig:P(r)}, we plot $P(\m{r}_\ell)$ for various $|U|/t$'s 
along a typical path on the lattice. 
Since $P({\bf r})$ is substantially constant for $|{\bf r}|\ge 2$ 
for any value of $|U|/t$, consistently with a QMC study,\cite{Guerrero} 
it is appropriate to put $P({\bf r})$ with the most distant 
${\bf r}=(L/2,L/2)$ at $P_\infty$, and check its system-size dependence. 
In Fig.~\ref{fig:P_infty_size}, we plot the thus-estimated $P_\infty$ 
as a function of $1/L$ for three values of $|U|/t$. 
In the SC state, the system-size dependence of $P_{\infty}$ is very weak 
for any $|U|/t$, and fitted well by a first-order least-squares method. 
Thus, we may discuss $P_\infty$ with a finite but large $L$. 
\begin{figure}[htbp]
\begin{center}
\includegraphics[width=7.0cm,clip]{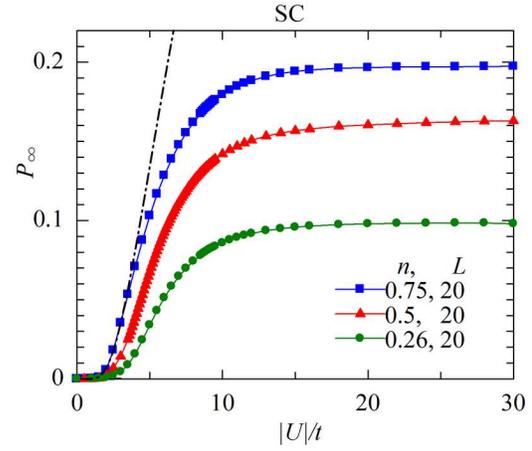}
\end{center}
\vspace{-0.5cm} 
\caption{(Color online) 
Onsite superconducting correlation function of $\Psi_{\rm SC}$ 
plotted as a function of $|U|/t$ for three particle densities 
and $L=20$. 
The dash-dotted line is a visual guide of $\propto\exp(-t/|U|)$ for $n=0.75$. 
}
\label{fig:P_infty}
\end{figure}
Figure \ref{fig:P_infty} shows the $|U|/t$ dependence of $P_\infty$ 
for $L=20$. 
In the BCS regime ($|U|\lsim|U_{\rm co}|$), $P_\infty$ increases as 
$\sim\exp(-t/U)$ as $|U|/t$ increases, as predicted by the BCS theory. 
In this regime, the optimized $\Delta_{\rm P}$ is considered to 
approximate the gap, as will be argued; we confirmed using the data 
shown in Fig.~\ref{fig:delta_var} that the relation 
$P_\infty\sim\Delta_{\rm P}^2$ holds. 
On the other hand, in the BEC regime, $P_\infty$ tends to converge 
at a finite value as $|U|/t$ increases, in accordance with the result 
of $\Delta_{\rm SC}$ obtained using DMFT.\cite{Garg,Bauer} 
Thus, the behavior of $P_\infty$ or $\Delta_{\rm SC}$ in this regime 
does not coincide with the behavior of $T_{\rm c}$, $\sim t^2/|U|$, 
which was naturally expected\cite{N-SR} and actually obtained by 
DMFT.\cite{Keller,Toschi,Koga} 
\par 

Incidentally, a similar SC correlation function with the NN $d_{x^2-y^2}$-wave 
pairing $P_d^\infty$ has been calculated for RHM in 2D by VMC with 
the same class of trial functions.\cite{Y12} 
In the strongly correlated regime (typically $U/t=12$), where the cuprates 
are considered to be properly described, $P_d^\infty$ behaves 
as the so-called dome shape as a function of doping rate $\delta$ ($=1-n$).  
This dome shape closely agrees with the $\delta$ dependence of $T_{\rm c}$ 
experimentally observed for the cuprates. 
If the framework of BSC-BEC crossover as a function of $\delta$ is 
applicable to this case, $P_d^\infty$ decreases and scales with $T_{\rm c}$
as the parameter approaches the BEC limit ($\delta\rightarrow 0$). 
Consequently, the behavior of $P_d^\infty$ in RHM does not fully 
correspond to that of $P_\infty$ in this study. 

\begin{figure}[htbp]
\begin{center}
\includegraphics[width=7.0cm,clip]{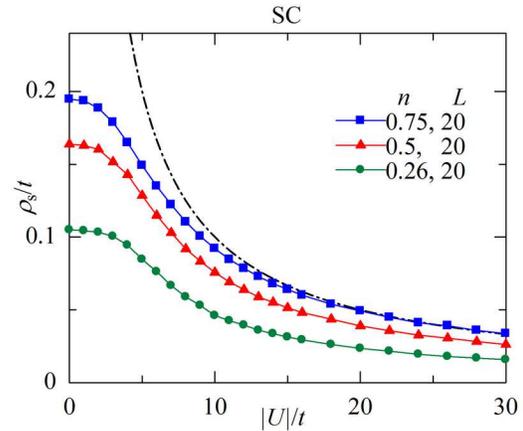}
\end{center}
\caption{(Color online) 
Helicity modulus as a function of $|U|/t$ for three particle densities. 
The dash-dotted line indicates a visual guide of $\propto -t/|U|$ for $n=0.75$. 
}
\vspace{-0.3cm} 
\label{fig:helicity}
\end{figure}
Now, we turn to the helicity modulus $\rho_{\rm s}$, which is related to 
superfluid stiffness $D_{\rm s}$ with $\rho_{\rm s}=D_{\rm s}/4\pi e^2$. 
In the SC state, $D_{\rm s}$ is equivalent to the strength of the 
delta-function component at $\omega=0$ in the optical conductivity 
$\sigma_1(\omega)$\cite{Scalapino} 
and represents the superfluid weight. 
We calculate $\rho_{\rm s}$ as the coefficient of the quadratic term 
in the increment in energy when the phase of order parameter 
$\Delta_j$ is twisted by ${\bf q}$ as 
$\Delta_j=|\Delta|e^{i{\bf q}\cdot{\bf r}_j}$:\cite{Fisher} 
\begin{equation}
E({\bf q})-E(0)=2\rho_{\rm s}{\bf q}^2+{\cal O}({\bf q}^4), 
\end{equation}
following the prescription of Denteneer \etal\cite{Denteneer} 
In Fig.~\ref{fig:helicity}, the $|U|/t$ dependence of the $\rho_{\rm s}$ thus 
obtained is plotted for three particle densities. 
The resultant $\rho_{\rm s}$ here is almost independent of the {\bf q} used 
($|{\bf q}|\sim 0.1$), and the system size dependence is negligible within 
the symbols between $L=12$ and 20. 
Regardless of $n$, $\rho_{\rm s}$ is a monotonically decreasing function 
of $|U|/t$. 
The behavior in the BCS regime is distinct from that of 
$T_{\rm c}\sim \exp(-t/|U|)$, but $\rho_{\rm s}$ scales with 
$T_{\rm c}\sim t^2/|U|$ in the BEC regime, indicating the strength 
of SC in the BEC regime is determined not by the cost of creating 
a pair but by the cost of realizing phase coherence. 
The present result of $\rho_{\rm s}$ is consistent with the previous 
results obtained by a VMC method with 
${\cal P}_{\rm G}\Phi_{\rm BCS}$,\cite{Denteneer} QMC,\cite{Singer} 
and DMFT.\cite{Garg,Toschi,Bauer}
\par 

\begin{figure}[htbp]
\begin{center}
\includegraphics[width=7.0cm,clip]{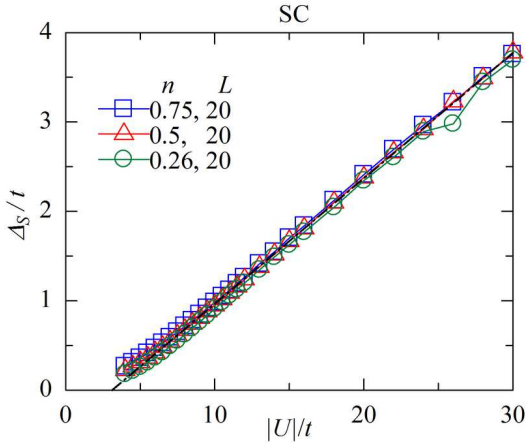}
\end{center}
\caption{(Color online) 
Spin gap of SC state estimated similarly to that in Fig.~\ref{fig:spingap} 
by single-mode approximation.
The dash-dotted straight line indicates an extrapolation from 
large-$|U|/t$ values for $n=0.5$. 
}
\vspace{-0.2cm} 
\label{fig:spingap_sc} 
\end{figure}
In the remainder of this subsection, we discuss some quantities related 
to $P({\bf r})$ and $\rho_{\rm s}$. 
First, we take up the small-$|{\bf q}|$ behavior of spin and charge 
structure factors, eqs.~(\ref{eq:Nq}) and (\ref{eq:Sq}). 
Like for the normal state, $N({\bf q})\propto |{\bf q}|$ 
for $|{\bf q}|\rightarrow 0$ for any $|U|/t$, showing that 
$\Psi_{\rm SC}$ is conductive in particle density. 
On the other hand, $S({\bf q})\propto |{\bf q}|^2$ for any $|U|/t$ ($>0$)  
in contrast to the case of $\Psi_{\rm N}$, indicating that a spin gap opens 
owing to pair formation.
We estimate the spin gap $\Delta_S$ for $\Psi_{\rm SC}$ using SMA 
[eq.~(\ref{eq:SMA})], and show the $|U|/t$ dependence 
in Fig.~\ref{fig:spingap_sc}. 
Although we do not display the data for small $|U|/t$'s owing to the 
relatively large errors due to the use of a finite system ($L=20$), $\Delta_S$ 
seems to be proportional to $\exp(-\alpha t/|U|)$ for small $|U|/t$'s. 
On the other hand, $\Delta_S$ is proportional to $|U|/t$, and has 
a magnitude similar to that of $\Psi_{\rm N}$. 
$\Delta_S$ is almost independent of $n$. 
Thus, $\Delta_S$ is a quantity that scales with $T_{\rm c}$ 
in the BCS regime. 
\par 

\begin{figure}[htbp]
\begin{center}
\includegraphics[width=7.0cm,clip]{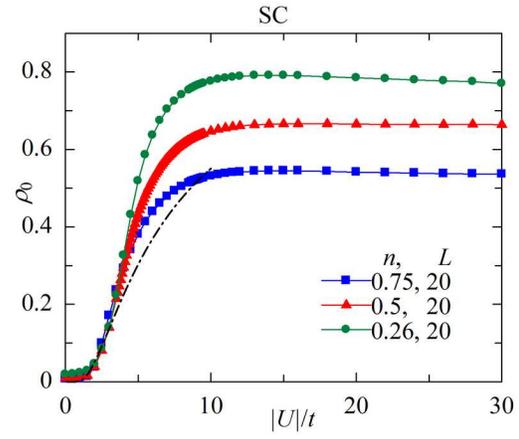}
\end{center}
\caption{(Color online) 
Condensate fractions of hard core bosons (or doublons) for three 
particle densities as functions of $|U|/t$. 
The dash-dotted line is a visual guide of $\propto \exp(-t/U)$ for n=0.26.
}
\vspace{-0.5cm} 
\label{fig:BEC1}
\end{figure}
Next, let us consider the condensate fraction $\rho_0$. 
We may regard $b^\dag_j$ as a creation operator of a hard-core 
spinless boson at the site $j$; $b^\dag_j$ satisfies the Bose 
commutation relation except for the same site. 
Following Bose systems,\cite{boson} we define a quantity corresponding to 
the condensate fraction or the ${\bf k}={\bf 0}$ element of momentum 
distribution function $n_D({\bf k})$ for $b_j^\dag$:  
\begin{equation}
 \rho_0=\frac{1}{D}n_D({\bf 0})= \frac{1}{DN_{\rm s}}
     \sum_{j,\ell}\langle b_{j}^{\dag}b_{j+\ell} \rangle.   
\label{eq:condensate}
\end{equation}
In Fig.~\ref{fig:BEC1}, the $|U|/t$ dependence of $\rho_0$ is depicted 
for three particle densities. 
The behavior of $\rho_0$ for small $|U|/t$'s is $\rho_0\sim\exp(-t/|U|)$ 
and has a meaning similar to $P_\infty$. 
In the BEC regime ($|U|>|U_{\rm co}|$), $\rho_0$ is almost constant, 
indicating that a picture of hard-core bosons is justified in 
the entire regime of BEC. 
The suppression of $\rho_0$ with increasing $n$ is primarily because 
a high density enhances the effect of onsite interaction. 
\par

Finally, we consider the pairing gap parameter $\Delta_{\rm P}$ given 
in eq.~(\ref{eq:ak}). 
In Fig.~\ref{fig:delta_var}, we show the $|U|/t$ dependence of the 
optimized $\Delta_{\rm P}$. 
For $|U|<|U|_{\rm co}$, it is natural that $\Delta_{\rm P}$ represents 
the SC gap of $\propto\exp(-t/|U|)$, as expected from the BCS theory. 
In the BCS theory, $\Delta_{\rm SC}$ should continue to 
linearly increase like $\Delta_S$ in Fig.~\ref{fig:spingap_sc}. 
However, the $\Delta_{\rm P}$ of $\Psi_{\rm SC}$ exhibits a peak at 
$U\sim U_{\rm co}$, and then decreases as $|U|/t$ increases, 
similarly to $\Delta E$ [Fig.~\ref{fig:cond_ene}(a)]. 
In the BEC regime, it is probable that $\Delta_{\rm P}$ obeys 
$\propto 2t^2/|U|=J$, because $J$ is the sole energy scale 
for $|U|/t\rightarrow\infty$, according to eq.~(\ref{eq:ste1}). 
It seems that $\Delta E$ and $\Delta_{\rm P}$ scale $T_{\rm c}$ in AHM. 
\par 

In the inset of Fig.~\ref{fig:delta_var}, we compare the optimized 
$\Delta_{\rm P}$'s between $\Psi_{\rm SC}$ and 
${\cal P}_{\rm G}\Phi_{\rm BCS}$. 
Although the two $\Delta_{\rm P}$'s behave similarly in the BCS regime, 
the $\Delta_{\rm P}$ of ${\cal P}_{\rm G}\Phi_{\rm BCS}$ monotonically 
increases unlike the $\Delta_{\rm P}$ of $\Psi_{\rm SC}$ in the BEC regime. 
It follows that the binding correlation between antiparallel spinons 
is also significant for $\Psi_{\rm SC}$, especially in the BEC regime. 
\par 

In variational theories with $d_{x^2-y^2}$-wave SC states for 
cuprates, the $d_{x^2-y^2}$-wave gap parameter $\Delta_d$, corresponding 
to $\Delta_{\rm P}$ here, is considered to represent a 
singlet-pairing gap (not necessarily SC gap).\cite{ZGRS,Paramekanti} 
The optimized $\Delta_d$ monotonically increases as the doping rate, 
the relevant parameter of the crossover, approaches the BEC limit 
($\delta\rightarrow 0$), in contrast 
to $T_{\rm c}$.\cite{ZGRS,Paramekanti,Y12} 
The behavior of $\Delta_{\rm p}$ here is distinct from that of $\Delta_d$. 
Again, we should be careful to consider the cuprate in the point of 
view of the BCS-BEC crossover. 
\par 

\subsection{Coherence length and intuitive picture
\label{sec:coherence}}
The BCS-BEC crossover has been typically explained by whether 
or not a domain of a Cooper pair overlaps with a domain of another 
pair, as in Fig.~\ref{fig:up-down}. 
To discuss this more quantitatively, we need to estimate a pair size 
$\xi_{\rm pair}$ corresponding to the coherence length 
and a distance between Cooper pairs $\tilde\xi_{\rm uu}$. 
As for $\xi_{\rm pair}$, it is reasonable to refer to the BCS expression 
of Pippard's coherence length, 
\begin{equation}
\xi_0=\frac{\hbar v_{\rm F}}{\pi|\Delta_{\rm SC}|}, 
\label{eq:xi-BCS}
\end{equation}
in the BCS regime. 
In the present study, $v_{\rm F}$ is a constant for any $|U|/t$, 
because the renormalization of $k_{\rm F}$ by $|U|/t$ is not 
introduced. 
Thus, we assume $\xi_{\rm pair}=\alpha/\Delta_{\rm P}$, where $\alpha$ is 
a constant determined so that $\xi_{\rm pair}$ can be smoothly connected 
to the form on the BEC side. 
In the BEC regime, eq.~(\ref{eq:xi-BCS}) does not work, because 
$v_{\rm F}$ cannot be defined. 
Thus, following eq.~(\ref{eq:xiud}), we naively assume 
$\xi_{\rm pair}=\tilde\xi_{\rm ud}=\tilde r_{\rm ud}+\tilde\sigma_{\rm ud}$, 
where $\tilde r_{\rm ud}$ and $\tilde\sigma_{\rm ud}$ denote 
the average distance between a spin (not necessarily of a spinon) 
and its nearest antiparallel spin and the standard deviation of 
$\tilde r_{\rm ud}$, respectively. 
Note that $\tilde r_{\rm ud}$ ($\tilde\sigma_{\rm ud}$) is different 
from $r_{\rm ud}$ ($\sigma_{\rm ud}$) in eq.~(\ref{eq:xiud}) in that 
spins that constitute doublons are taken into account.  
Similarly, following eq.~(\ref{eq:xiuu}), we estimate an interpair 
distance as the average minimum distance between a spin and 
its nearest parallel spin, 
$\tilde\xi_{\rm uu}=\tilde r_{\rm uu}-\tilde\sigma_{\rm uu}$. 
\par

\begin{figure}[htbp]
\begin{center}
\includegraphics[width=7.0cm,clip]{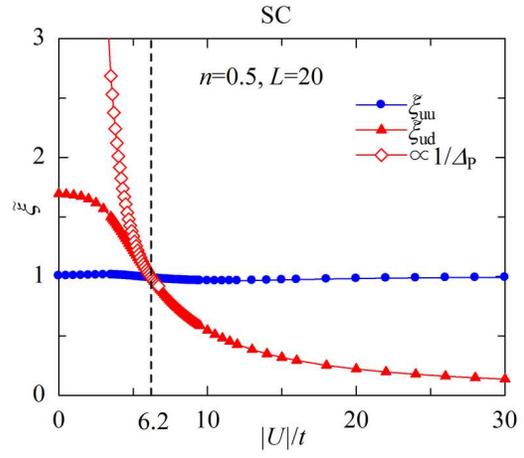}
\end{center}
\caption{(Color online) 
Two estimates ($\alpha/\Delta_{\rm P}$ and $\tilde\xi_{\rm ud}$) of 
the pair size $\xi_{\rm pair}$, corresponding to the coherence length, 
and the average minimum distance between Cooper pairs $\tilde\xi_{\rm uu}$ 
are compared as functions of $|U|/t$. 
The dashed line indicates $|U|/t$ where $\xi_{\rm pair}$ 
intersects $\tilde\xi_{\rm uu}$. 
}
\label{fig:xi_sc}
\end{figure}
In Fig.~\ref{fig:xi_sc}, the $|U|/t$ dependences of $\xi_{\rm pair}$ and 
$\tilde\xi_{\rm uu}$ thus estimated are compared for $n=0.5$. 
The pair size $\xi_{\rm pair}$ is a monotonically decreasing 
function of $|U|/t$, whereas the interpair distance $\tilde\xi_{\rm uu}$ 
is almost independent of $|U|/t$. 
Consequently, $\xi_{\rm pair}$ crosses $\tilde\xi_{\rm uu}$ at 
$|U|=|U_\xi|\sim 6.2t$ at this particle concentration. 
Thus, for $|U|<|U_\xi|$, Cooper pairs penetrate each other 
($\xi_{\rm pair}>\tilde\xi_{\rm uu}$) as in the Fermi liquid phase 
in Fig.~\ref{fig:up-down}. 
On the other hand, for $|U|>|U_\xi|$, a pair becomes almost a point 
hard-core boson ($\xi_{\rm pair}\sim 0$), and is isolated from other pairs 
($\xi_{\rm pair}<\tilde\xi_{\rm uu}$). 
Similarly, we estimated $|U_\xi|/t$ for other values of $n$, and 
found $|U_\xi|\sim 6t$ irrespective of particle densities.\cite{note-xi} 
Since the above estimation of $U_\xi$ is rather broad, we consider that
$U_\xi$ should be identical to $U_{\rm co}$. 
\par

Finally, we point out the difference in pairing manner between the 
spin-gap transition at $U_{\rm c}$ in the normal state (\S\ref{sec:picture}) 
and the crossover at $\sim U_{\rm co}$ in the SC state. 
Bound spinon pairs in the normal state ($|U|>|U_{\rm c}|$) dissociate 
into independent spinons immediately when the pair domains overlap 
with each other at $U=U_{\rm c}$. 
On the other hand, Cooper pairs in the SC state remain paired, even if 
pair domains come to considerably overlap for $|U|\ll |U_{\rm co}|$ 
in the BCS regime; there is no critical change at $U=U_{\rm co}$. 
Consequently, a phase transition (crossover) arises and a spin gap 
closes (survives) on the weakly correlated side in the normal 
(SC) state. 
The stability of Cooper pairs against mutual overlap is a current 
topic.\cite{Zhu} 
\par

\section{Conclusions\label{sec:summary}}
Using a variational Monte Carlo (VMC) method, we studied the features of 
a spin-gap transition in a normal state and of the BCS-BEC crossover 
in a superconducting (SC) state in the attractive Hubbard model (AHM) 
on the square lattice. 
We summarize the main results below.
\par

(1) 
In the normal state, we revealed that, unlike the simple Gutzwiller
wave function (GWF), a wave function with an antiparallel-spinon binding 
correlation $P_Q$ [eqs.~(\ref{eq:PQ}) and (\ref{eq:Qj})] undergoes 
a first-order transition from a Fermi liquid to a spin-gapped phase 
at $|U_{\rm c}|/t\sim 9$. 
In the spin-gapped phase, particle density current can flow through 
the hopping of doublons. 
The pseudogap phase above $T_{\rm c}$ for $|U|\gsim |U_{\rm c}|$ may be 
deduced from the properties of this wave function.
The mechanism of this spin-gap transition is understood to be similar to that 
of a Mott transition in a repulsive Hubbard model (RHM) induced by 
a doublon-holon binding correlation.\cite{Miyagawa,boson}  
We would also like to realize a variational normal state that is spin-gapped 
and conductive in RHM.
\par

(2) 
We first applied VMC to the SC state of AHM, and confirmed that, 
as $|U|/t$ increases, the mechanism of superconductivity undergoes 
a crossover at approximately $|U_{\rm co}|\sim|U_{\rm c}|$ from an BCS type 
to a Bose-Einstein condensation (BEC) type. 
$P_Q$ is again needed to suppress the gap, which is greatly overestimated 
in GWF for $|U|\gsim |U_{\rm co}|$. 
In the weak-correlation regime ($|U|<|U_{\rm co}|$), the strength of SC 
($T_{\rm c}$) is scaled with quantities related to the SC gap
as $\sim\exp(-t/|U|)$, as expected from the BCS theory. 
For $|U|>|U_{\rm co}|$, the superfluid stiffness, which is related to 
the cost of phase coherence, scales with $T_{\rm c}$ as $t^2/|U|$.   
Such typical features of this crossover are captured by the energy gain 
in the SC transition $\Delta E$ in the whole range of $|U|/t$. 
In the BEC regime, the SC transition is induced by a gain in kinetic 
energy; this aspect is in contrast to the BCS theory, but is in accord with 
the magnetic and SC transitions in strongly correlated RHM.\cite{YTOT,Y12} 
Most features are consistent with the framework of BCS-BEC crossover 
that previous studies provided.\par 

(3) 
The physics of a spin-gap transition in the normal state and the BCS-BEC 
crossover in the SC state are explained in a semiquantitative manner 
by a simple idea of the competition between the pair size $\xi_{\rm ud}$ 
and the interpair distance $\xi_{\rm uu}$, as shown 
in Fig.~\ref{fig:up-down}. 
This idea is equivalent to that of Mott transitions 
in RHM,\cite{Miyagawa,boson} in which a doublon-holon pair corresponds 
to the singlet pair here.
\par

(4) 
In connection with that observed high-$T_{\rm c}$ cuprates, the $|U|/t$ 
dependence of the pair correlation function $P_\infty$ and the gap parameter 
$\Delta_{\rm P}$ studied here qualitatively differ from the doping rate 
($\delta$) dependence of the corresponding quantities ($P_d^\infty$ and 
$\Delta_d$) in the strongly correlated RHM, when the relevant parameters 
($|U|/t$ and $\delta$) are in the respective BEC regimes. 
Furthermore, in a strongly correlated RHM, the $d_{x^2-y^2}$-wave SC 
transition is always kinetic-energy-driven, regardless of $n$.\cite{Y12} 
We will address this subject more carefully in upcoming publications. 
\par 

\bigskip
\begin{acknowledgments} 
\noindent
{\bf Acknowledgments}
\par\medskip
The authors thank Tomoaki Miyagawa for helpful discussions. 
This work is partially supported by Grant-in-Aids from the Ministry of 
Education, Culture, Sports, Science, and Technology, Japan. 
\end{acknowledgments}




\begin{thebibliography}{99}
\bibitem{note-Uc} In the DMFT result in ref.~\citen{Capone}, 
$|U_{\rm c}|/t$ abruptly decreases as $n\rightarrow 0$; this behavior 
quantitatively differs from the present one. 
We consider that this is partly because $\Psi_{\rm N}$ does not allow for 
long-range correlations, which becomes effective as $n$ approaches 
zero.
\bibitem{note-kappa} 
The compressibility $\kappa=(n^2\partial^2 E/\partial n^2)^{-1}$ of 
$\Psi_{\rm N}$ tends to be negative at high densities 
($0.7\lsim n \lsim 1.0$), whether the phase is a Fermi liquid or spin-gapped. 
We leave it for future studies to determine whether this instability against 
phase separation is intrinsic in the normal phase of AHM in 2D or 
an artifact of $\Psi_{\rm N}$. 
As for $\Psi_{\rm SC}$, $\kappa$ is always positive.
\bibitem{note-co} 
The principal feature of the BCS-BEC crossover is a change in the mechanism 
of SC transition, which arises owing to the competition between 
the SC and normal states. 
Therefore, the BCS-BEC crossover is not necessarily a problem of the SC 
state alone; some quantities such as $\Delta E$ may reflect the properties 
of the normal state.
\bibitem{noteP} 
Onsite pair correlation $P(\m{r}_\ell)$ is predominant at least for large 
$|U|/t$'s, but we leave the effect of longer-distance pairs 
for small $|U|/t$'s for future study.
\bibitem{note-xi} 
We checked whether the repulsive correlation factor ${\cal P}_f$ 
[eq.~(\ref{eq:pf})] between parallel spins markedly affects 
$\tilde\xi_{\rm uu}$. 
The resultant $\tilde\xi_{\rm uu}$ by ${\cal P}_{\cal F}\Phi_{\rm BCS}$ 
for $n=0.26$ is only 1\% longer than the value here.
\end{thebibliography}
\end{document}